\documentclass[%
 aip,jcp,
 amsmath,amssymb,
 reprint,
nolongbibliography 
]{revtex4-2}

\usepackage{graphicx}
\usepackage{dcolumn}
\usepackage{bm}

\usepackage[utf8]{inputenc}
\usepackage[T1]{fontenc}
\usepackage{mathptmx}
\usepackage{listings}
\usepackage{rotating} 

\usepackage{color}
\usepackage{dcolumn} 
\usepackage{bm} 
\usepackage{graphicx}
\usepackage{multirow} 
\usepackage{pifont} 
\usepackage{epsfig}
\usepackage{amsmath} 
\usepackage{subfigure}
\usepackage{float}
\usepackage{booktabs}
\usepackage{tabularx}
\usepackage{natbib}
\usepackage{gensymb}
\usepackage{rotating}
\usepackage{threeparttable}
\usepackage{comment}
\usepackage{physics}
\usepackage[normalem]{ulem}
\usepackage[hidelinks]{hyperref} 

\begin{document}
  
\author{Stephen H. Yuwono}
\affiliation{
             Department of Chemistry and Biochemistry,
             Florida State University,
             Tallahassee, FL 32306-4390}          

\author{Brandon C. Cooper}
\affiliation{
             Department of Chemistry and Biochemistry,
             Florida State University,
             Tallahassee, FL 32306-4390}       

\author{Tianyuan Zhang}
\affiliation{Department of Chemistry, University of Washington, Seattle, WA 98195}

\author{Xiaosong Li}
\affiliation{Department of Chemistry, University of Washington, Seattle, WA 98195}
             
\author{A. Eugene DePrince III}
\email{adeprince@fsu.edu}
\affiliation{
             Department of Chemistry and Biochemistry,
             Florida State University,
             Tallahassee, FL 32306-4390}

\title{Time-Dependent Equation-of-Motion Coupled-Cluster Simulations with a Defective Hamiltonian}

\begin{abstract}

Simulations of laser-induced electron dynamics in a molecular system are performed using time-dependent (TD) equation-of-motion (EOM) coupled-cluster (CC) theory. The target system has been chosen to highlight potential shortcomings of truncated TD-EOM-CC methods [represented in this work by TD-EOM-CC with single and double excitations (TD-EOM-CCSD)], where unphysical spectroscopic features can emerge. Specifically, we explore driven resonant electronic excitations in magnesium fluoride in the proximity of an avoided crossing. Near the avoided crossing, the CCSD similarity-transformed Hamiltonian is defective, meaning that it has complex eigenvalues{\color{black}, and oscillator strengths may take on negative values.} When an external field is applied to drive transitions to states exhibiting these traits, unphysical dynamics are observed. For example, the stationary states that make up the time-dependent state acquire populations that can be negative, exceed one, or even be complex-valued. 

\end{abstract}

\maketitle

\section{Introduction}
\label{SEC:INTRODUCTION}

Recent years have seen increased interest in the use of real-time (RT) time-dependent (TD) electronic structure methodologies\cite{Li20_review,Pedersen23_e1666} to describe electron dynamics or to directly simulate various spectroscopic techniques, such as UV/visible spectroscopy,\cite{Cramer15_1102,DePrince16_5834,Sheng18_1800055} X-ray absorption and photoemission spectroscopy,\cite{DePrince17_2951,Bartlett19_164117,Peng20_174113,Peng20_6983,Bartlett21_094103,Rehr22_1799,Repisky23_1714} circular dichroism,\cite{Li16_234102,Li19_6824,DePrince19_204107} and pump-probe / transient absorption spectroscopy.\cite{Mukamel12_194306,Govind15_4294,Rubio16_128,Parkhill16_1590,Lopata20_4470,Koch22_023103,Repisky23_1714} Among the plethora of RT-TD methods that have been put forward to model such phenomena, a large number of studies have employed RT-TD formulations of density functional theory (DFT).\cite{Gross84_997,Theilhaber92_12990,Bertsch96_4484} The popularity of RT-TD-DFT is not surprising, given its low computational cost and oftentimes reasonable accuracy. However, as has been demonstrated numerous times,\cite{Nest12_806,Isborn14_184112,Isborn15_4791,Dewhurst22_062409} RT-TD-DFT cannot properly describe electron dynamics of systems driven far from the ground state; within the adiabatic approximation, the positions of peaks shift as the relative populations of states change, which makes it impossible to drive the system between two states with a resonant external field. This deficiency precludes the direct application of RT-TD-DFT to simulations of time-resolved spectroscopy experiments.

Given the failures of RT-TD-DFT, one could turn to wave-function-based \emph{ab initio} approaches that incorporate electronic correlation effects beyond the mean-field approximation. Among the methods in this category, those based on coupled-cluster (CC)\cite{Coester58_421,Kuemmel60_477,Cizek66_4256,Cizek69_35,Shavitt72_50,Li99_1,Musial07_291} theory and its extensions to excited states using the equation-of-motion (EOM)\cite{Emrich81_379,Bartlett89_57,Bartlett93_7029} formalism show particular promise, as they have the potential to  accurately model both linear-response absorption spectra and long-time, driven electron dynamics. The strengths of TD-CC and TD-EOM-CC follow from the desirable properties of their time-independent counterparts, including the size extensivity and separability of truncated CC approaches, the size intensivity of EOM-CC excitation energies, and the rapid convergence of the CC/EOM-CC energetics toward the exact, full configuration interaction (CI), limit.\cite{Musial07_291} To illustrate the last point, the familiar CC with single and double excitations (CCSD) approximation\cite{Bartlett82_1910,Zerner77_4088} and its excited-state analogue (EOM-CCSD)\cite{Bartlett93_7029} offer qualitatively correct descriptions of many chemical problems of interest, while higher-order methods that also incorporate triple excitations (CCSDT/EOM-CCSDT),\cite{Bartlett87_7041,Schaefer88_382,Bartlett90_6104,Piecuch01_643,Piecuch01_237,Bartlett01_8263} or triple and quadruple excitations (CCSDTQ/EOM-CCSDTQ),\cite{Adamowicz91_6645,Bartlett91_387,Bartlett92_4282,Adamowicz94_5792,Gauss04_9257,Hirata04_51} practically recover the full CI description of the system, albeit with substantially increased computational costs.\cite{Musial07_291}

Multiple routes toward time-dependent CC/EOM-CC methods have been proposed. The most direct strategy is to incorporate time-dependence into the cluster amplitudes,
as is done in the TD-CC approach.\cite{Klamroth11_054113}
However, the original formulation of TD-CC utilized static HF orbitals, which introduces instabilities into the {\em ansatz} when the reference configuration has a small overlap with the time-dependent state.\cite{Kvaal19_144106,Pedersen20_071102,Koch23_arxiv2301.05546}
{\color{black}On the other hand, the orbital-adaptive TD-CC (OA-TD-CC)\cite{Kvaal12_194109} approach, which was originally developed as a gauge invariant formulation of TD-CC, makes use of a set of time-dependent, biorthogonal orbitals. Treating the orbitals on the same dynamical footing as the cluster amplitudes improves the stability of the approach, but OA-TD-CC still presents numerical instabilities in the presence of strong electric fields.\cite{Pedersen20_071102} 
The TD orbital-optimized CC (TD-OCC) approach\cite{Ishikawa18_051101,Ishikawa20_124115,Ishikawa20_034110,Ishikawa21_234104} uses a single set of time-dependent orthogonal orbitals and is more stable than OA-TD-CC at high field strengths, but it exhibits other problems, chief among which is the fact that TD-OCC does not converge to the TD full CI limit.\cite{Olsen05_084116}}
An alternative to TD-CC{\color{black}, OA-TD-CC, and TD-OCC} is to fix the cluster amplitudes {\color{black}and orbitals} at their ground-state values and incorporate time-dependence through the EOM-CC excitation operators, as is done in the TD-EOM-CC approach.\cite{Schlegel11_4678,Head-gordon12_909,DePrince16_5834,DePrince17_2951,DePrince19_204107,Li19_6617,Bartlett19_164117,Bartlett21_094103,DePrince21_5438,Koch22_023103,Koch23_arxiv2301.05546} The TD-EOM-CC propagation has been shown to be more stable relative to {\color{black} TD-CC when} the system is driven far from the ground state.\cite{Koch23_arxiv2301.05546}
Furthermore, the linear nature of TD-EOM-CC is desirable from an implementation standpoint, whereas TD-CC{\color{black}, OA-TD-CC, and TD-OCC} involve the time-evolution of more complicated systems of non-linear equations.

One aspect of the TD-CC/EOM-CC that has not been explored, to the best of our knowledge, is the effect that the non-Hermitian nature of the similarity-transformed Hamiltonian can have on the time-evolution of the CC/EOM-CC wave function. It is well known that the similarity-transformed Hamiltonian of a truncated EOM-CC method may {\color{black}exhibit} complex energy eigenvalues, which are clearly unphysical, in the neighborhood of a conical intersection.{\color{black} \cite{Hattig05_37,Koch17_164105,Gauss21_e1968056}} {\color{black}In particular, truncated EOM-CC can fail to describe conical intersections between states of the same symmetry when the relevant diagonal elements of the similarity-transformed Hamiltonian matrix are small and the off-diagonal elements coupling the states have opposite signs. In these cases, the similarity-transformed Hamiltonian matrix can become defective, which means that it is impossible to find non-parallel eigenvectors that diagonalize it without moving into the complex plane. We would} like to understand the behavior of TD-EOM-CC in regions of space at or near such {\color{black}defects.}
In this study, we focus on TD-EOM-CCSD simulations of MgF, which exhibits complex eigenvalues due to the failure of EOM-CCSD to properly describe an avoided crossing between two $\Sigma^+$ symmetry states. In addition, near the regions where the energy is complex, some excited states exhibit negative oscillator strengths. As we will demonstrate, several peculiar behaviors emerge when the system is driven to these unphysical states.

This paper is organized as follows. In Sec.\ \ref{SEC:THEORY}, we provide a brief summary of the TD-EOM-CC {\color{black} formalism.}
The details of the calculations performed in this work are then outlined in Sec.\ \ref{SEC:COMPUTATIONAL_DETAILS}. In Sec.\ \ref{SEC:RESULTS}, we discuss the results of our computations for MgF, where we highlight the emergence of the unphysical properties mentioned above (complex eigenvalues and negative oscillator strengths), which result from expanding the similarity-transformed Hamiltonian in an incomplete many-electron basis. We then compare the results of TD-EOM-CCSD simulations performed in regions where the similarity-transformed Hamiltonian displays these unphysical characteristics versus where EOM-CCSD energies and properties appear to be well-behaved.
{\color{black} We also briefly discuss the effects that symmetrization of the CCSD similarity-transformed Hamiltonian has on the energies and oscillator strengths.}
Finally, we conclude our discussion in Sec.\ \ref{SEC:CONCLUSIONS}.

\section{Theory}
\label{SEC:THEORY}

Within the TD-EOM-CC framework, one generates the initial state ({\em i.e.}, the state at time $t = 0$) by solving the stationary CC ground-state problem. In doing so, we employ the wave function {\em ansatz}
\begin{equation}
\label{EQN:CC_GS}
    \ket*{\Psi_0} = \exp(\hat{T}) \ket{\Phi_0}
\end{equation}
where $\ket{\Phi_0}$ represents a Hartree--Fock determinant, and $\hat{T}$ is the cluster operator, which can be expressed as 
\begin{equation}
\label{EQN:CLUSTER}
     \hat{T} = \sum_{n=1}^{N} \left(\frac{1}{n!}\right)^{2}
    t_{a_1 \ldots a_n}^{i_1 \ldots i_n} \hat{E}_{i_1 \ldots i_n}^{a_1 \ldots a_n}
\end{equation}
Here, $t_{a_1 \ldots a_n}^{i_1 \ldots i_n}$ is a cluster amplitude, and $\hat{E}_{i_1 \ldots i_n}^{a_1 \ldots a_n} = \Pi_{\lambda=1}^{n} \hat{a}^{a_\lambda} \hat{a}_{i_\lambda}$ is the elementary $n$-body hole--particle excitation operator. The symbols $\hat{a}_{p}$ and $\hat{a}^{p}\equiv\hat{a}_{p}^\dagger$ represent the annihilation and creation operators corresponding to the spin orbital $\phi_p$. Throughout this paper, we use the Einstein summation convention, where repeated upper and lower indices are summed, and the indices $i_1,i_2,\ldots$ and $a_1,a_2,\ldots$ refer to occupied and unoccupied spin orbitals, respectively.

The cluster amplitudes are determined by solving the energy-independent projection equations
\begin{equation}
\label{EQN:CC_PROJ}
    \mel*{\Phi_{i_1\ldots i_n}^{a_1\ldots a_n}}{\bar{H}}{\Phi_0} = 0\quad \forall n = 1,\ldots,N
\end{equation}
where $\bar{H} = \exp(-\hat{T}) \hat{H} \exp(\hat{T})$ is the similarity-transformed electronic Hamiltonian. The ground-state CC energy is then computed as the expectation value of $\bar{H}$ with respect to the reference determinant, \emph{i.e.},
\begin{equation}
\label{EQN:CC_GS_EN}
    E_0 = \mel{\Phi_0}{\bar{H}}{\Phi_0}
\end{equation}
Up until this point, we have not made any approximations, so the resulting ground-state energy $E_0$ in Eq.\ \ref{EQN:CC_GS_EN} is equivalent to that obtained from an exact full CI computation. However, full CC calculations with realistic basis sets and more than a few electrons are prohibitively expensive, so, in practice, the cluster expansion in Eq.\ \ref{EQN:CLUSTER} and the corresponding manifold of excited Slater determinants defining the projections in Eq.\ \ref{EQN:CC_PROJ} are usually truncated at an order lower than $N$, resulting in the well-known hierarchy of CCSD, CCSDT, CCSDTQ, \emph{etc.}\ approaches.

Before moving on to the TD-EOM-CC formalism, let us briefly consider the time-independent EOM-CC framework. In EOM-CC, the $K$-th excited-state wave function ($K>0$) is parameterized as
\begin{equation}
\label{EQN:EOM_KET}
    \ket*{\Psi_K} = \hat{R}_K \exp(\hat{T}) \ket{\Phi_0}
\end{equation}
where, $\hat{R}_K$ is a linear excitation operator that can be expanded as
\begin{equation}
\label{EQN:EOM_R}
    \hat{R}_K = r_{K,0}\mathbf{1} + \sum_{n=1}^{N}  \left(\frac{1}{n!}\right)^{2} r_{K,a_1 \ldots a_n}^{\phantom{K,}i_1 \ldots i_n} \hat{E}_{i_1 \ldots i_n}^{a_1 \ldots a_n}
\end{equation}
with
{\color{black} $r_{K,0}$ being the weight of the ground-state wave function} and
$r_{K,a_1 \ldots a_n}^{\phantom{K,}i_1 \ldots i_n}$ being the EOM excitation amplitudes.
{\color{black} By inserting Eq.~\ref{EQN:EOM_KET} as the wave function in the time-independent Schr\"odinger equation [$\hat{H}|\Psi_K\rangle = E_K|\Psi_K\rangle$], multiplying from the left by $\exp(-\hat{T})$,
and subtracting $E_0 \mathbf{1} \hat{R}_K \ket{\Phi_0}$ from both sides of the equation,} one arrives {\color{black}[after taking into account that $\hat{R}_K$ and $\exp(\hat{T})$ commute]} at the eigenvalue equation
\begin{equation}
\label{EQN:EOM_EIGENVALUE}
    \bar{H}_N \hat{R}_K \ket{\Phi_0} = \omega_K \hat{R}_K \ket{\Phi_0}
\end{equation}
{\color{black}Here,} $\bar{H}_N = \bar{H} - E_0\mathbf{1}$ is the normal-ordered form of the similarity-transformed Hamiltonian and $\omega_K = E_K - E_0$ is the vertical excitation energy corresponding to the $K$-th excited state. It is immediately apparent that one can solve for the EOM excitation amplitudes and the excitation energies by diagonalizing $\bar{H}_N$. However, one must keep in mind that the similarity-transformed Hamiltonian is not Hermitian, and, consequently, if one is interested in properties other than the energy, one should also consider the left-hand or ``bra'' EOM-CC eigenvectors
\begin{equation}
\label{EQN:EOM_BRA}
    \bra*{\tilde{\Psi}_K} = \bra{\Phi_0} \hat{L}_K \exp(-\hat{T})
\end{equation}
In this case, $\hat{L}_K$ is a linear \emph{de-excitation} operator
\begin{equation}
\label{EQN:EOM_L}
    \hat{L}_K = \delta_{K0}\mathbf{1} + \sum_{n=1}^{N}  \left(\frac{1}{n!}\right)^{2} l_{K,i_1 \ldots i_n}^{\phantom{K,}a_1 \ldots a_n}
    (\hat{E}_{i_1 \ldots i_n}^{a_1 \ldots a_n})^\dagger
\end{equation}
where $\delta_{K0}$ is a Kronecker delta and $l_{K,i_1 \ldots i_n}^{\phantom{K,}a_1 \ldots a_n}$ are the $n$-body EOM de-excitation amplitudes.
{\color{black} Inserting Eq.\ \ref{EQN:EOM_BRA} into the complex-conjugate form of Schr{\"o}dinger equation, multiplying from the right by $\exp(\hat{T})$, and subtracting $\bra{\Phi_0} \hat{L}_K \mathbf{1} E_0$ from both sides of the equation produces the left-hand counterpart of Eq.\ \ref{EQN:EOM_EIGENVALUE},
\begin{equation}
\label{EQN:EOM_EIGENVALUE_L}
    \bra{\Phi_0} \hat{L}_K \bar{H}_N  = \bra{\Phi_0} \hat{L}_K \omega_K
\end{equation}
}The left and right EOM-CC eigenvectors satisfy the relationship
\begin{equation}
\label{EQN:EOM_BIORTHONORMAL}
    \braket*{\tilde{\Psi}_J}{\Psi_K} = 
    \mel*{\Phi_0}{\hat{L}_J \hat{R}_K}{\Phi_0} = \delta_{JK}
\end{equation}
and, thus, form a biorthonormal basis for $\bar{H}$
{\color{black} under the assumption that the matrix $\bar{H}$ is diagonalizable.}
To compute a property of interest, one only needs to evaluate the quantity
\begin{equation}
\label{EQN:EOM_PROPERTY}
    \mel*{\tilde{\Psi}_J}{\hat{O}}{\Psi_K} =
    \mel*{\Phi_0}{\hat{L}_J \bar{O} \hat{R}_K}{\Phi_0}
\end{equation}
where $\bar{O} = \exp(-\hat{T})\hat{O}\exp(\hat{T})$. In analogy to the ground-state case, if $\hat{R}_K$ and $\hat{L}_K$ include all components up to the ones that excite/de-excite $N$ electrons, EOM-CC recovers the corresponding full CI energetics and other properties of the $K$-th state. In practice, one truncates $\hat{R}_K$ and $\hat{L}_K$ at the same level as the cluster operator, and $\bar{H}$ is diagonalized in the truncated space (or partially diagonalized via iterative methods such as the Davidson procedure to obtain a subset of excited states). 
As noted in {\color{black} Refs.\ \onlinecite{Koch17_4801,Koch17_164105,Gauss21_e1968056}}, and as we show below, truncated EOM-CC calculations may produce complex eigenvalues under certain situations, which is not too surprising, given the fact that $\bar{H}$ is not Hermitian.

We now move on to the main focus of this work, which is the time-evolution of EOM-CC wave functions in the presence of an external perturbation. In this case, the Hamiltonian is time-dependent and, within the dipole approximation, takes the form
\begin{equation}
\label{EQN:TD_HAMILTONIAN}
    \hat{H}(t) = \hat{H} - \hat{\mu}\cdot \hat{\varepsilon}(t)
\end{equation}
where $\hat{\mu}$ is the dipole moment operator, and $\hat{\varepsilon}(t)$ is the applied external electric field.
We generalize the  EOM left- and right-hand states (Eqs.\ \ref{EQN:EOM_KET} and \ref{EQN:EOM_BRA}), to be time-dependent as well, as
\begin{equation}
\label{EQN:TD_EOM_KET}
    \ket{\Psi(t)} = \hat{R}(t) \exp(\hat{T}) \ket{\Phi_0}
\end{equation}
and
\begin{equation}
\label{EQN:TD_EOM_BRA}
    \bra{\tilde{\Psi}(t)} = \bra{\Phi_0} \hat{L}(t) \exp(-\hat{T})
\end{equation}
The TD-EOM operators entering Eqs.\ \ref{EQN:TD_EOM_KET} and \ref{EQN:TD_EOM_BRA} are now
\begin{equation}
\label{EQN:TD_EOM_R}
    \hat{R}(t) = r_0(t)\mathbf{1} + \sum_{n=1}^{N} \left(\frac{1}{n!}\right)^{2} r_{a_1 \ldots a_n}^{i_1 \ldots i_n}(t) \hat{E}_{i_1 \ldots i_n}^{a_1 \ldots a_n}
\end{equation}
and
\begin{equation}
\label{EQN:TD_EOM_L}
    \hat{L}(t) = l_0(t)\mathbf{1} + \sum_{n=1}^{N}  \left(\frac{1}{n!}\right)^{2} l_{i_1 \ldots i_n}^{a_1 \ldots a_n}(t)
    (\hat{E}_{i_1 \ldots i_n}^{a_1 \ldots a_n})^\dagger
\end{equation}
with the time-dependence of the wave function encoded in the $r$ and $l$ amplitudes. The initial conditions at $t=0$ are  $\hat{R}(0) = 1$ and $\hat{L}(0) = 1 + \sum_{n=1}^{N} \left(\frac{1}{n!}\right)^{2} \lambda_{i_1 \ldots i_n}^{a_1 \ldots a_n} (\hat{E}_{i_1 \ldots i_n}^{a_1 \ldots a_n})^\dagger$, where the $\lambda$ amplitudes are the solution of the ground-state left-hand CC equations.
{\color{black} We proceed by inserting the right- and left-hand TD-EOM-CC wave functions into the TD Schr{\"o}dinger equation
\begin{equation}
\label{EQN:TDSE_ket}
    i \dv{\ket{\Psi(t)}}{t} = \hat{H}(t)\ket{\Psi(t)}
\end{equation}
and its complex-conjugate form, respectively. By recognizing that the time-derivative operator acts only on $\hat{R}(t)$ and $\hat{L}(t)$, and not on the exponentiated cluster operators nor on the reference determinant, and using steps similar to those used to arrive at Eqs.\ \ref{EQN:EOM_EIGENVALUE} and \ref{EQN:EOM_EIGENVALUE_L}, we obtain the TD-EOM-CC Schr{\"o}dinger equation
}
\begin{equation}
\label{EQN:TDSE_R}
    i \dv{\hat{R}(t)}{t}\ket{\Phi_0} = \bar{H}(t)\hat{R}(t)\ket{\Phi_0}
\end{equation}
and its complex-conjugate
\begin{equation}
\label{EQN:TDSE_L}
    -i \bra{\Phi_0}\dv{\hat{L}(t)}{t} = \bra{\Phi_0}\hat{L}(t)\bar{H}(t)
\end{equation}
{\color{black}Here,} we have introduced the similarity-transformed TD Hamiltonian $\bar{H}(t) = \bar{H} - \bar{\mu}\vdot\hat{\epsilon}(t)${\color{black}, where $\bar{\mu} = {\rm exp}(-\hat{T})\hat{\mu}{\rm exp}(\hat{T})$ is the similarity-transformed dipole operator.} In order to evaluate time-dependent observables, such as the dipole response function, one simply computes the appropriate expectation value as a function of time. For the dipole response, we have
\begin{equation}
\label{EQN:TD_DIPOLE}
    \expval{\hat{\mu}(t)} = \mel*{\Phi_0}{\hat{L}(t) \bar{\mu} \hat{R}(t)}{\Phi_0}
\end{equation}

We can simulate different optical phenomena with different choices for the functional form of the external field. For example, consider a Gaussian envelope centered at $t_0>0$
\begin{equation}
\label{EQN:GAUSS_SINE_FIELD}
    \hat{\varepsilon}(t) = \va{\varepsilon}_\mathrm{max} \exp(\frac{(t-t_0)^2}{2\sigma^2}) \sin(\omega_\mathrm{ex}t)
\end{equation}
where $\va{\varepsilon}_\mathrm{max}$ is a vector defining the field strength and polarization, $\sigma$ is the width of the Gaussian, and $\omega_\mathrm{ex}$ is the carrier frequency of the pulse. By inserting this form of the field into Eqs.\ \ref{EQN:TDSE_R} and \ref{EQN:TDSE_L} and performing Fourier transform of the resulting signal defined in Eq.\ \ref{EQN:TD_DIPOLE}, one obtains the linear absorption spectrum
\begin{equation}
\label{EQN:FT_LINESHAPE}
    f(\omega) = \frac{2}{3} \omega \;\Im\left[\frac{\int_{-\infty}^{\infty} e^{i\omega t - \gamma t} \expval{\hat{\mu}(t)} \dd t}{\int_{-\infty}^{\infty} e^{i\omega t} \hat{\varepsilon}(t) \dd t}\right]
\end{equation}
where we have introduced a damping parameter $\gamma$ that corresponds to the half-width at half-maximum of the peaks in a Lorentzian-broadened spectrum. Note also that the Fourier transform of the dipole moment that enters the oscillator strength expression is also normalized by the Fourier transform of the external field. On the other hand, with a different choice for $\hat{\varepsilon}(t)$, we can simulate driven resonant dynamics ({\em i.e.}, Rabi oscillations). In this case, we choose a continuous wave of the form
\begin{equation}
\label{EQN:SINE_FIELD}
    \hat{\varepsilon}(t) = \va{\varepsilon}_\mathrm{max} \sin(\omega_\mathrm{ex}t)
\end{equation}
and the resulting generalized Rabi frequency is
\begin{equation}
\label{EQN:RABI_FREQ} 
    \Omega_{0K} = \sqrt{(\va{\varepsilon}_\mathrm{max}\vdot \mu_{0K})( \va{\varepsilon}_\mathrm{max}\vdot\mu_{K0})  + \left\vert\omega_\mathrm{ex} - \omega_{K}\right\vert^2}
\end{equation}
where 
\begin{align}
    \mu_{JK} = \mel{\tilde{\Psi}_J}{\hat{\mu}}{\Psi_K}
\end{align}

\section{Computational Details}
\label{SEC:COMPUTATIONAL_DETAILS}

In order to examine how the unphysical characteristics of truncated similarity-transformed Hamiltonian and the corresponding EOM-CC eigenvectors affect the electron dynamics driven by an external electric field, we performed frequency- and time-domain EOM-CCSD calculations for the MgF molecule using the STO-3G basis set.\cite{Pople69_2657,Pople70_2769} We have calculated CCSD and EOM-CCSD energies, dipole and transition dipole moments, and oscillator strengths of MgF at inter-atomic distances ($R_\mathrm{Mg\mbox{-}F}$) of  $1.600$--2.000 \AA, at 0.001 \AA\ increments. After identifying bright states with complex eigenvalues and/or negative oscillator strength values, we performed TD-EOM-CCSD simulations of driven transitions to these states at selected points in the $R_\mathrm{Mg\mbox{-}F} = 1.600$--2.000 \AA\ region.

The frequency-domain CCSD/EOM-CCSD calculations and subsequent TD-EOM-CCSD propagations reported in this work were performed using a Python code developed with the help of the \texttt{p$^\dagger$q} second quantized algebra package.\cite{DePrince21_e1954709} This code was interfaced with the unrestricted Hartree--Fock (UHF) and integral transformation routines of \textsc{Psi4}.\cite{Sherrill20_184108} The UHF calculations were carried out using $C_{2v}$ symmetry. Given the small size of the MgF/STO-3G system (21 electrons in 28 spin orbitals), we were able to form the full CCSD $\bar{H}$ in the space of the reference and singly and doubly substituted determinants with $S_z=+\tfrac{1}{2}$ and diagonalize it, allowing us access to the full spectrum of the CCSD $\bar{H}$. Furthermore, $\bar{H}$ was blocked according to irreducible representations of the $C_{2v}$ point group to prevent arbitrary rotations among degenerate states. The occurrence of complex eigenvalues and negative or complex oscillator strength values was verified against EOM-CCSD data computed using generalized HF reference in the Chronus Quantum software package.\cite{Li19_e1436} We have also performed additional computations based on the restricted open-shell HF (ROHF) reference, as well as using $C_1$ point group, and we have verified that the unphysical behaviors persist in these cases (see Supplementary Material).
{\color{black} In order to investigate the effects of basis set size on the occurrence of complex eigenvalues, we also performed additional CCSD/EOM-CCSD calculations using the 6-31G\cite{Pople72_2257,Hehre1982_2797,Pople1982_3654}, cc-pVDZ,\cite{Dunning89_1007,Wilson11_69} and cc-pVTZ\cite{Dunning89_1007,Wilson11_69} basis sets. Because our Python implementation is not efficient enough to construct the full CCSD $\bar{H}$ with these basis sets, we computed the lowest 40 excited states of MgF with $A_1$ symmetry at $R_\mathrm{Mg\mbox{-}F}=1.70$--2.30 \AA\ with 0.01 \AA\ step size in \textsc{Psi4}, making sure that the geometry and energy ranges encompass those of the complex seam observed in the STO-3G case (see below).}

Our TD-EOM-CCSD propagations were carried out in the $\bar{H}$ eigenbasis representation, where $\bar{H}$ is a diagonal matrix [although the $\bar{\mu}\vdot\hat{\varepsilon}(t)$ term still introduces non-Hermiticity in the overall structure of $\bar{H}(t)$], and the left and right eigenstates are unit vectors. This choice allows us to express the TD Schr\"odinger equations as simple matrix-vector multiplications and provides us with easy access to population analysis, which is useful when analyzing the Rabi cycle simulations. Equations \ref{EQN:TDSE_R} and \ref{EQN:TDSE_L} were integrated using the {\color{black} explicit} 4th-order Runge--Kutta scheme, with a time step $\Delta t = 0.01$ a.u.\ ($\sim2.4\times10^{-19}$ s) and a total simulation time of $t = 5000$ a.u. To produce the linear absorption spectrum of MgF, we employed Eq.\ \ref{EQN:GAUSS_SINE_FIELD}  with $\vert\va{\varepsilon}_\mathrm{max}\vert =10^{-5}$ a.u., $t_0 = 10.0$ a.u., $\sigma=7.07\times 10^{-3}$ a.u., and $\omega_\mathrm{ex} = 1.0$ a.u., and the Fourier transform to obtain the frequency-domain spectrum was performed with $\gamma=0.005$ a.u.\ (0.136 eV). In our simulation of laser-driven excitations using Eq.\ \ref{EQN:SINE_FIELD}, we used $\vert\va{\varepsilon}_\mathrm{max}\vert = 2.0\times10^{-3}$ a.u. We stress-tested our code by performing TD-EOM-CCSD propagations with different field strengths, time steps, and total simulation times, and we found that above parameters produce stable TD propagations within the time ranges in which we are interested. {\color{black} Lastly, we also monitored the probability density $\braket{\tilde{\Psi}(t)}{\Psi(t)}$, which we found to be conserved at unity throughout all simulations. Some of the} data produced in the stability tests can be found in the Supplementary Material.

\section{Results and Discussion}
\label{SEC:RESULTS}

\subsection{Frequency-domain EOM-CCSD data}

\begin{figure}[!htbp]
    \centering
    \includegraphics[width=0.5\textwidth]{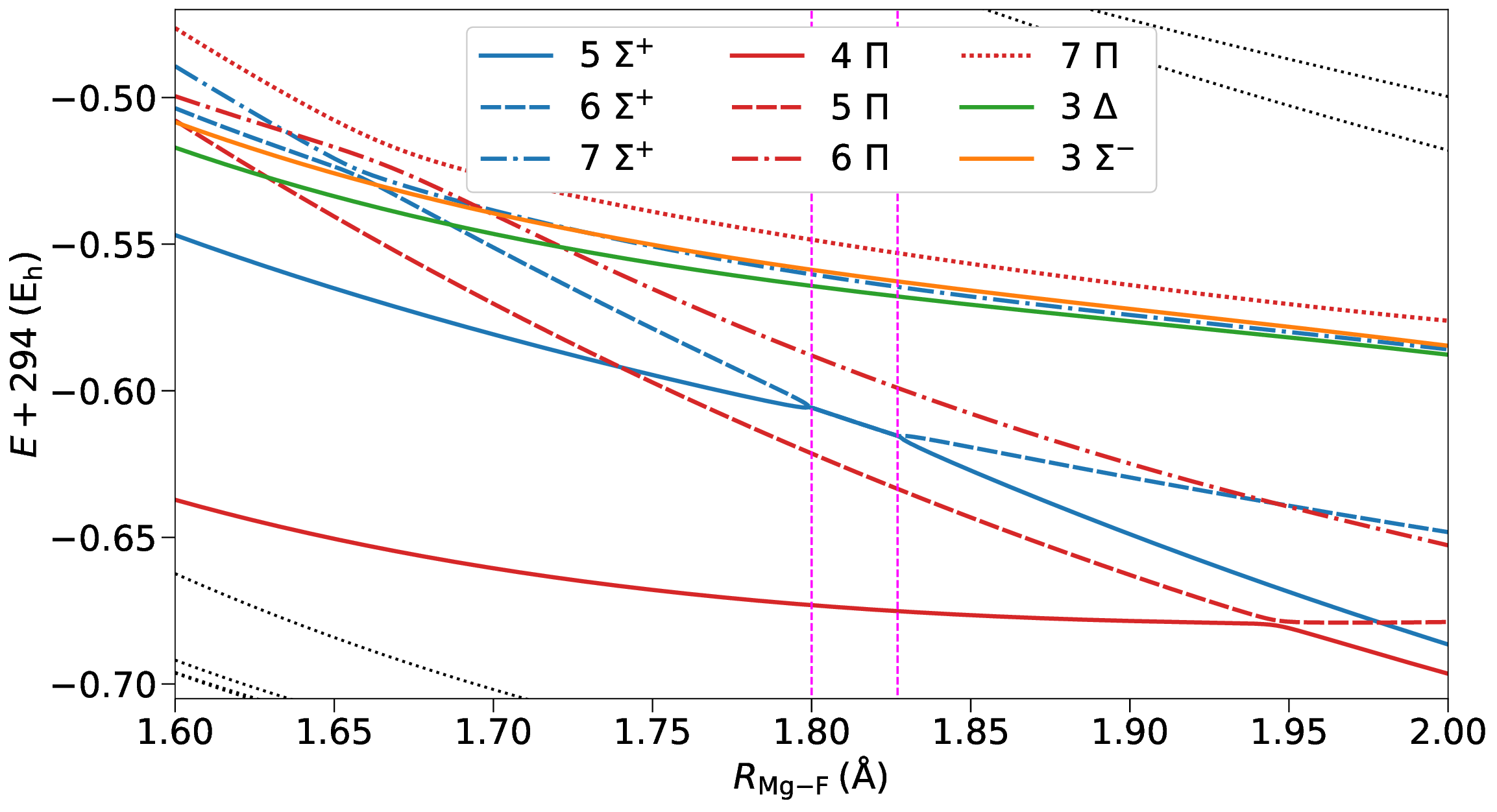}
    \caption{Potential energy curves of selected excited states of MgF computed at the EOM-CCSD/STO-3G level of theory. The states are labeled according to increasing energy at $R_\mathrm{Mg\mbox{-}F} = R_\mathrm{e} = 1.692$ \AA. Note the occurrence of complex eigenvalues resulting from the failure of EOM-CCSD to properly describe the avoided crossing between the $5\;\Sigma^{+}$ and $6\;\Sigma^{+}$ states
    ({\color{black}blue solid and dashed lines}, respectively) in the [1.800,1.827] \AA\
    {\color{black} region, which is demarcated with vertical magenta lines. Other states that are energetically far from the relevant states are marked with black dotted lines.}
    }
    \label{fig:MgF_PEC}
\end{figure}

\begin{figure}[!htbp]
    \centering
    \includegraphics[width=0.5\textwidth]{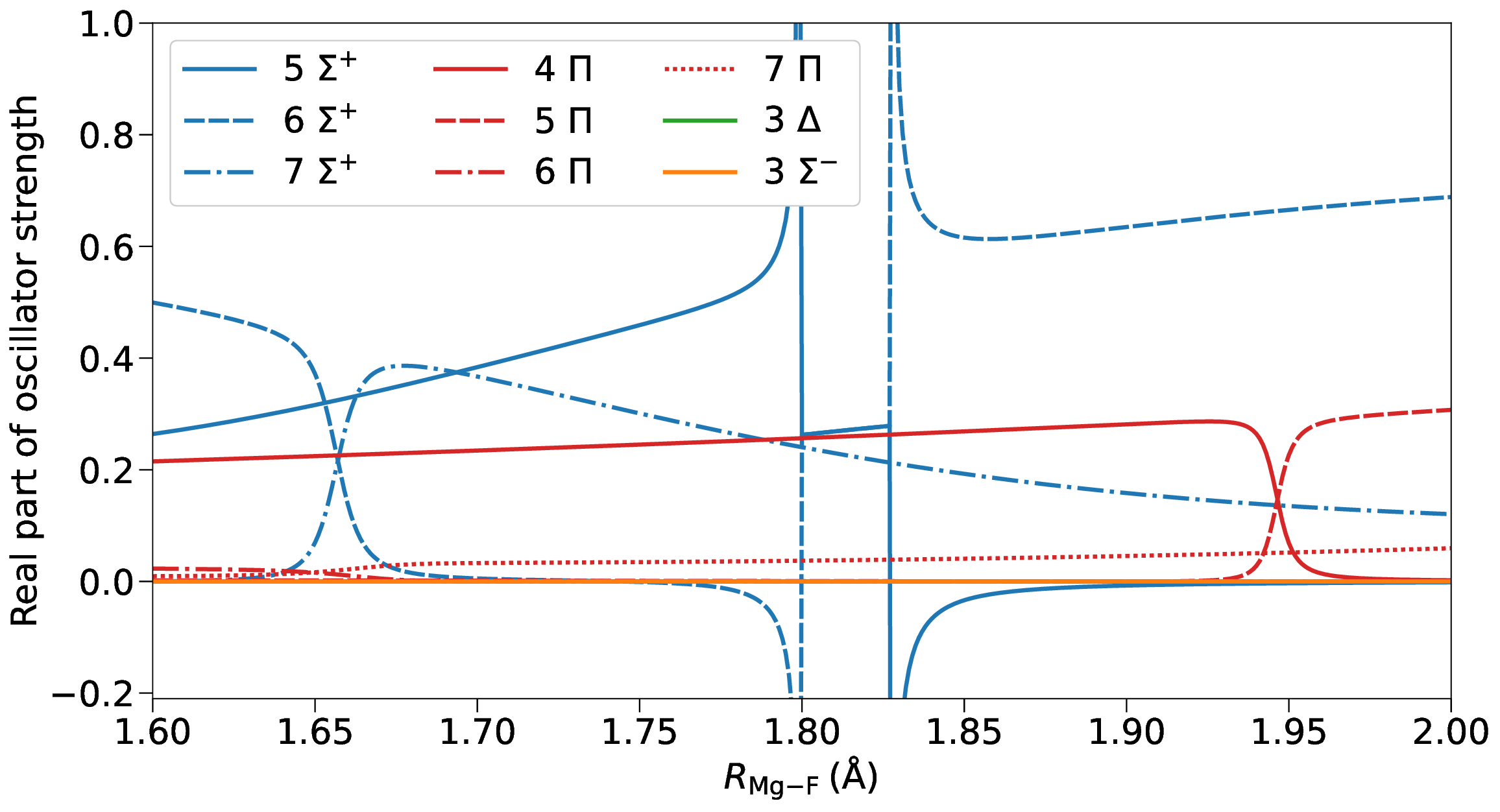}
    \caption{Oscillator strength curves (real part) of selected excited states of MgF computed at the the EOM-CCSD/STO-3G level of theory. The states are labeled according to increasing energy at $R_\mathrm{Mg\mbox{-}F} = R_\mathrm{e} = 1.692$ \AA. Note the occurrence of complex oscillator strengths resulting from the failure of EOM-CCSD to properly describe the avoided crossing between the $5\;\Sigma^{+}$ and $6\;\Sigma^{+}$ states
    ({\color{black}blue solid and dashed lines}, respectively) in the [1.800,1.827] \AA\ region.}
    \label{fig:MgF_OS}
\end{figure}

We begin by analyzing the results of frequency-domain EOM-CCSD/STO-3G calculations of MgF in the $R_\mathrm{Mg\mbox{-}F} = 1.600$--2.000 \AA\ region, which are summarized in Figs.\ \ref{fig:MgF_PEC} and \ref{fig:MgF_OS} and Tables \ref{tab:MgF_omega_f_sigmaplus} and \ref{tab:MgF_omega_f_othersym}. Figure \ref{fig:MgF_PEC} depicts the potential energy curves (PECs) of several excited states of MgF in the proximity of the region where the energies of the 5th and 6th $\Sigma^+$ states become complex (between $R_\mathrm{Mg\mbox{-}F} = 1.800$ and 1.827 \AA), where it appears that EOM-CCSD fails to correctly describe an avoided crossing between these states. This failure manifests in properties other than energy, such as the oscillator strength ($f$) curves, which are reported in Fig.\ \ref{fig:MgF_OS}. We see that EOM-CCSD is able to properly describe other avoided crossings in this region, \emph{e.g.}, the avoided crossings between the $6\;\Sigma^{+}$ and $7\;\Sigma^{+}$ states near $R_\mathrm{Mg\mbox{-}F} = 1.65$ \AA\ and between the $4\;\Pi$ and $5\;\Pi$ states at around 1.95 \AA; in these cases, Fig.~\ref{fig:MgF_OS} shows the clear exchange of the oscillator strength character between the two curves at the avoided crossing. In contrast, the values of $f$ for the $5\;\Sigma^{+}$ and $6\;\Sigma^{+}$ states diverge as the Mg--F bond distance approaches the [1.800,1.827] \AA\ region, from either direction. Along the complex seam, the oscillator strength values, like the energy, become complex.
While the failure of EOM-CCSD to properly describe the avoided crossing  results in a somewhat recognizable feature of the PEC, the divergence of the oscillator strengths suggests that EOM-CCSD wave functions of the $5\;\Sigma^{+}$ and $6\;\Sigma^{+}$ states have dramatically different properties near the complex seam than far away from it. 

\begin{figure}[htbp]
    \centering
    \includegraphics[width=0.5\textwidth]{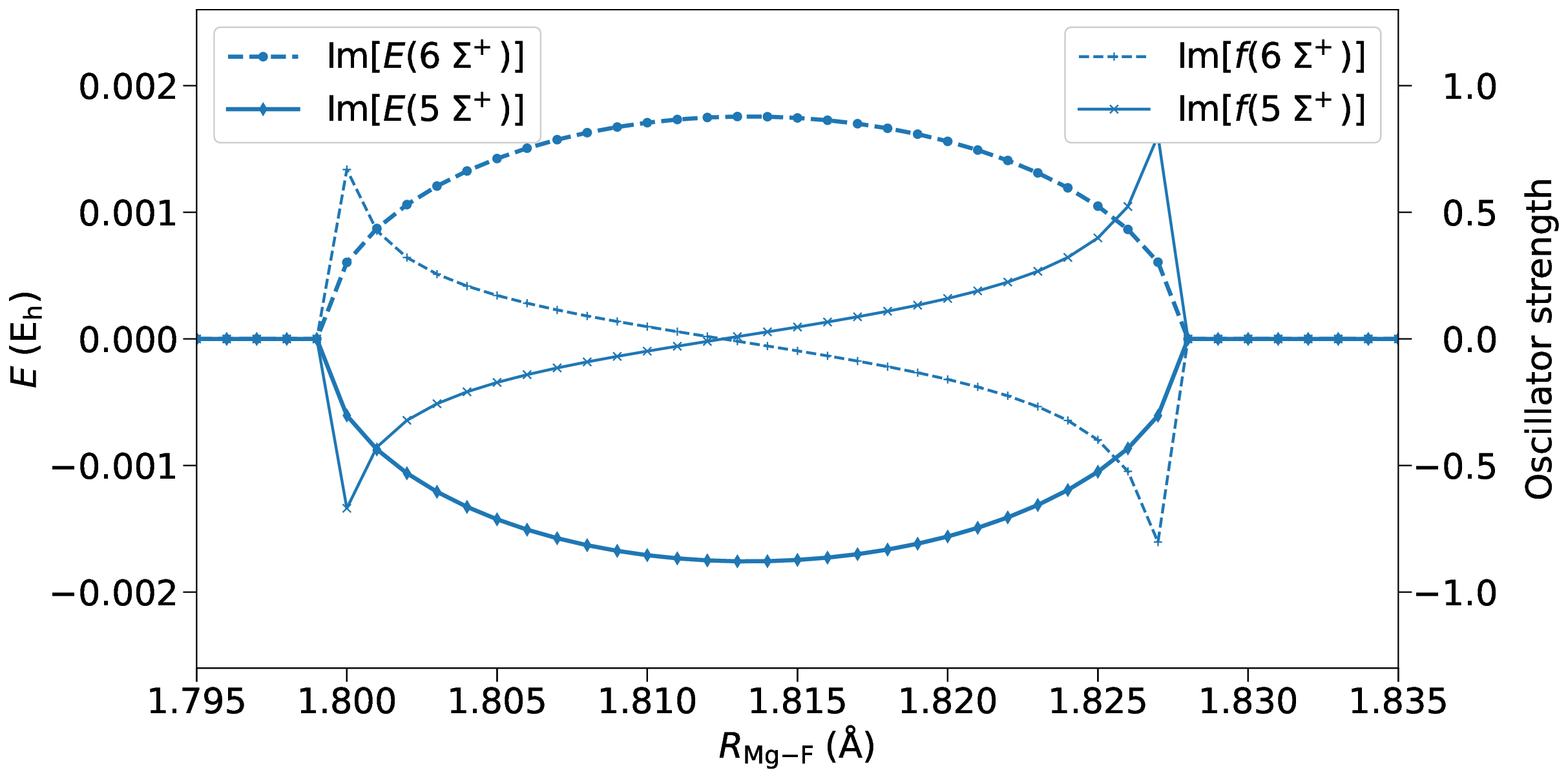}
    \caption{The imaginary components of the energy and oscillator strength values characterizing the $5\;\Sigma^{+}$ {\color{black} (blue solid lines)}
    and $6\;\Sigma^{+}$ {\color{black} (blue dashed lines)}
    states of MgF in the $R_\mathrm{Mg\mbox{-}F} = 1.795$--1.835 \AA\ region.}
    \label{fig:MgF_imaginary}
\end{figure}

For selected bond distance values, the energies and oscillator strengths of the states shown in Figs.\ \ref{fig:MgF_PEC} and \ref{fig:MgF_OS}  are reported in Tables \ref{tab:MgF_omega_f_sigmaplus} and \ref{tab:MgF_omega_f_othersym}. Focusing on Table \ref{tab:MgF_omega_f_sigmaplus}, we highlight two interesting observations related to the imaginary components of the energy and oscillator strength values. First, at $R_\mathrm{Mg\mbox{-}F}=1.800$ \AA, which is the shortest bond length at which we observe complex eigenvalues in Fig.\ \ref{fig:MgF_PEC}, the imaginary components of the oscillator strengths for the $5\;\Sigma^{+}$ and $6\;\Sigma^{+}$ states are much larger in magnitude than the real parts. On the other hand, the imaginary components of the energy eigenvalues are small in comparison to the real parts. This observation could reflect the contrasting behavior of the real parts of the energies and oscillator strength values of the $5\;\Sigma^{+}$ and $6\;\Sigma^{+}$ states in Figs.\ \ref{fig:MgF_PEC} and \ref{fig:MgF_OS} as one approaches the complex seam, {\em i.e.}, the energies converge, while the oscillator strength values diverge. Second, for the $5\;\Sigma^{+}$ and $6\;\Sigma^{+}$ states, the signs of the imaginary components of the energy do not change as we pass through the complex seam, but those of the oscillator strengths do. This behavior is illustrated in Fig.\ \ref{fig:MgF_imaginary}. 
As we can see, there is a transfer of character between the states when the sign on the imaginary components flips between $R_\mathrm{Mg\mbox{-}F} = 1.812$ and 1.813 \AA. These observations are consistent with the argument that the complex seam represents a failed attempt of EOM-CCSD to describe an avoided crossing; the avoided crossing has tilted onto the imaginary axis, which is where we also observe the exchange of character of the oscillator strengths. Furthermore, the data in Fig.~\ref{fig:MgF_imaginary} also show that, while the imaginary energy components smoothly approach zero as the Mg--F bond length gets nearer to either end of the complex seam, as we return to the well-behaved geometries, the oscillator strengths diverge; this divergence is similar to that observed for the real component of the oscillator strength values shown in Fig.~\ref{fig:MgF_OS}.

\begin{table*}[!htbp]
    \caption{\label{tab:MgF_omega_f_sigmaplus} The EOM-CCSD/STO-3G vertical excitation energies ($\omega$, in $\mathrm{E_h}$) and oscillator strengths ($f$) characterizing the excited states of MgF with $\Sigma^{+}$ symmetry close to the occurrence of complex eigenvalues at selected Mg--F bond distances (in \AA). The ground-state CCSD energies (in $\mathrm{E_h}$) are shown for reference.}
    
    \centering
    \begin{tabular}{c @{\extracolsep{8pt}}c @{\extracolsep{8pt}}cc @{\extracolsep{8pt}}cc @{\extracolsep{8pt}}cc}
        \hline\hline
        \multirow{2}{*}{$R_\mathrm{Mg\mbox{-}F}$} & \multirow{2}{*}{$E(1\;\Sigma^{+})$} & \multicolumn{2}{c}{$5\;\Sigma^{+}$} & \multicolumn{2}{c}{$6\;\Sigma^{+}$} & \multicolumn{2}{c}{$7\;\Sigma^{+}$} \\
        \cline{3-4} \cline{5-6} \cline{7-8}
        & & $\omega$ & $f$ & $\omega$ & $f$ & $\omega$ & $f$ \\
        \hline
        1.600 & $-295.131147$ & 0.5842           & 0.2640           & 0.6275           & 0.4995           & 0.6418 & 0.0000 \\
        1.692 & $-295.136994$ & 0.5586           & 0.3722           & 0.5904           & 0.0075           & 0.6008 & 0.3762 \\
        1.790 & $-295.132145$ & 0.5278           & 0.5650           & 0.5324           & $-0.0527$        & 0.5736 & 0.2515 \\
        1.800 & $-295.131203$ & $0.5255-0.0006i$ & $0.2625-0.6688i$ & $0.5255+0.0006i$ & $0.2625+0.6688i$ & 0.5709 & 0.2404 \\
        1.820 & $-295.129127$ & $0.5162-0.0016i$ & $0.2746+0.1596i$ & $0.5162+0.0016i$ & $0.2746-0.1596i$ & 0.5656 & 0.2196 \\
        1.830 & $-295.128002$ & 0.5106           & $-0.3275$        & 0.5126           & 0.8882           & 0.5630 & 0.2101 \\
        1.900 & $-295.118843$ & 0.4699           & $-0.0073$        & 0.4893           & 0.6349           & 0.5447 & 0.1582 \\
        2.000 & $-295.103627$ & 0.4171           & $-0.0016$        & 0.4554           & 0.6884           & 0.5178 & 0.1205 \\
        \hline\hline
    \end{tabular}
\end{table*}

\begin{table*}[!htbp]
    \caption{\label{tab:MgF_omega_f_othersym} The EOM-CCSD/STO-3G vertical excitation energies ($\omega$, in $\mathrm{E_h}$) and oscillator strengths ($f$) characterizing the excited states of MgF with $\Sigma^{-}$, $\Pi$, and $\Delta$ symmetries, close to the occurrence of complex eigenvalues at selected Mg--F bond distances (in \AA).}
    
    \centering
    \begin{tabular}{c @{\extracolsep{8pt}}cc @{\extracolsep{8pt}}cc @{\extracolsep{8pt}}cc @{\extracolsep{8pt}}cc @{\extracolsep{8pt}}cc @{\extracolsep{8pt}}cc}
        \hline\hline
        \multirow{2}{*}{$R_\mathrm{Mg\mbox{-}F}$} & \multicolumn{2}{c}{$3\;\Sigma^{-}$} & \multicolumn{2}{c}{$4\;\Pi$} & \multicolumn{2}{c}{$5\;\Pi$} & \multicolumn{2}{c}{$6\;\Pi$} & \multicolumn{2}{c}{$7\;\Pi$} & \multicolumn{2}{c}{$3\;\Delta$}\\
        \cline{2-3} \cline{4-5} \cline{6-7} \cline{8-9} \cline{10-11} \cline{12-13}
        & $\omega$ & $f$ & $\omega$ & $f$ & $\omega$ & $f$ & $\omega$ & $f$ & $\omega$ & $f$ & $\omega$ & $f$ \\
        \hline
        1.600 & 0.6227 & 0.0000 & 0.4940 & 0.2149 & 0.6233 & 0.0016 & 0.6316 & 0.0231 & 0.6548 & 0.0094 & 0.6141 & 0.0000 \\
        1.692 & 0.5995 & 0.0000 & 0.4779 & 0.2328 & 0.5712 & 0.0015 & 0.6013 & 0.0003 & 0.6120 & 0.0325 & 0.5923 & 0.0000 \\
        1.790 & 0.5750 & 0.0000 & 0.4599 & 0.2541 & 0.5154 & 0.0007 & 0.5485 & 0.0001 & 0.5854 & 0.0365 & 0.5693 & 0.0000 \\
        1.800 & 0.5725 & 0.0000 & 0.4581 & 0.2564 & 0.5098 & 0.0006 & 0.5432 & 0.0001 & 0.5827 & 0.0371 & 0.5670 & 0.0000 \\
        1.820 & 0.5674 & 0.0000 & 0.4545 & 0.2612 & 0.4987 & 0.0005 & 0.5329 & 0.0000 & 0.5772 & 0.0384 & 0.5622 & 0.0000 \\
        1.830 & 0.5649 & 0.0000 & 0.4527 & 0.2637 & 0.4933 & 0.0004 & 0.5278 & 0.0000 & 0.5745 & 0.0391 & 0.5598 & 0.0000 \\
        1.900 & 0.5468 & 0.0000 & 0.4403 & 0.2816 & 0.4560 & 0.0000 & 0.4940 & 0.0000 & 0.5549 & 0.0455 & 0.5426 & 0.0000 \\
        2.000 & 0.5190 & 0.0000 & 0.4071 & 0.0021 & 0.4248 & 0.3072 & 0.4509 & 0.0001 & 0.5275 & 0.0594 & 0.5159 & 0.0000 \\
        \hline\hline
    \end{tabular}
\end{table*}

{\color{black} At this point, it is worth asking whether the occurrence of complex eigenvalues in MgF may be related to the use of the minimal (STO-3G) basis set. As mentioned above, we also carried out PEC scans for MgF with the 6-31G, cc-pVDZ, and cc-pVTZ basis sets to shed light on this issue. The complex seam persists in the 6-31G case, but not in the cc-pVDZ or cc-pVTZ calculations (at least in the vicinity of the geometry and energy ranges observed in the STO-3G case). Thus, for this molecule, complex eigenvalues can be avoided by employing a sufficiently large basis set, without increasing the truncation order in $\hat{T}$, $\hat{R}$, or $\hat{L}$. On the other hand, complex eigenvalues have been observed in EOM-CCSD calculations on larger molecules and basis sets (see, for example, Refs.\ \onlinecite{Koch17_4801,Gauss21_e1968056}). Additional systematic studies of the effects of basis set size and the level of electron correlation treatment on the occurrence of complex eigenvalues could provide valuable insights.}

\subsection{Linear absorption spectra from TD-EOM-CCSD}

\begin{figure}[htbp]
    \centering
    \includegraphics[width=0.5\textwidth]{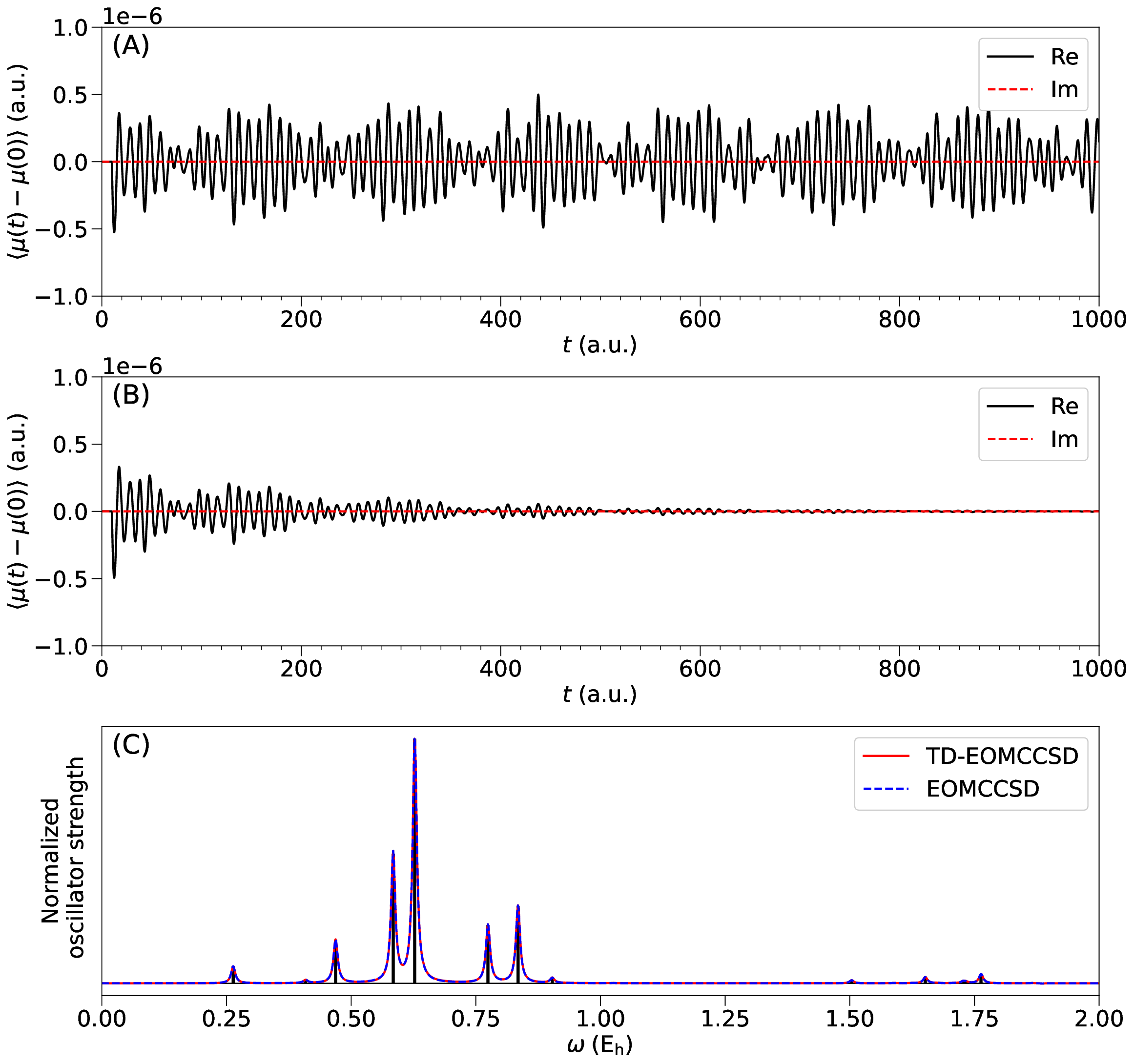}
    \caption{Dipole response and linear absorption spectrum of MgF at $R_\mathrm{Mg\mbox{-}F} = 1.600$ \AA\ with $z$-axis polarization. (A) The dipole response signal up to $t=1000$ a.u. (B) The dampened dipole response signal using $\gamma=0.005$ a.u.\ or 0.136 eV. (C) The linear absorption spectrum obtained from the Fourier transform of the dampened dipole response signal (for a total simulation time of  5000 a.u., red solid line) and Lorentzian convolution of frequency-domain EOM-CCSD oscillator strengths (blue dashed line). The stick spectrum from EOM-CCSD oscillator strengths is also included.}
    \label{fig:MgF_LR_1.6}
\end{figure}

\begin{figure}[htbp]
    \centering
    \includegraphics[width=0.5\textwidth]{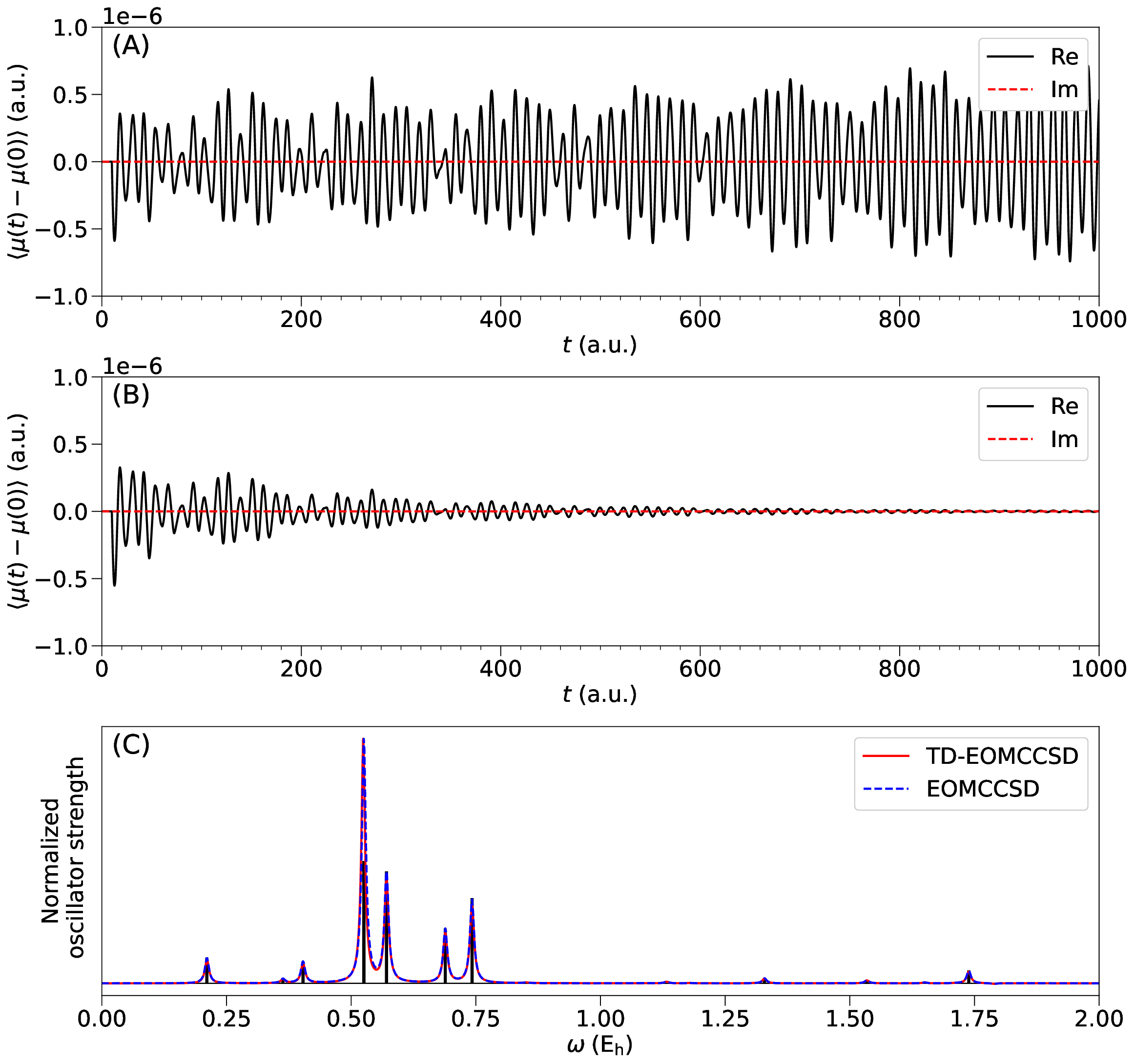}
    \caption{Dipole response and linear absorption spectrum of MgF at $R_\mathrm{Mg\mbox{-}F} = 1.800$ \AA\ with $z$-axis polarization. (A) The dipole response signal up to $t=1000$ a.u. (B) The dampened dipole response signal using $\gamma=0.005$ a.u.\ or 0.136 eV. (C) The linear absorption spectrum obtained from the Fourier transform of the dampened dipole response signal (for a total simulation time of  5000 a.u., red solid line) and Lorentzian convolution of frequency-domain EOM-CCSD oscillator strengths (blue dashed line). The stick spectrum from EOM-CCSD oscillator strengths is also included.} 
    \label{fig:MgF_LR_1.8}
\end{figure}

We now consider how the presence of the complex-valued eigenvalues of $\bar{H}$ impact TD-EOM-CCSD simulations of linear absorption spectra. Toward this aim, we have performed simulations at a ``well-behaved'' geometry (1.600 \AA, Fig.\ \ref{fig:MgF_LR_1.6}) and a geometry known to exhibit complex energy eigenvalues (1.800 \AA, Fig.\ \ref{fig:MgF_LR_1.8}), with the field polarized along the $z$ axis to simulate transitions to states with $\Sigma^{+}$ symmetry. The functional form for the field is given by Eq.~\ref{EQN:GAUSS_SINE_FIELD}. Note that, in Figs.\ \ref{fig:MgF_LR_1.6} and \ref{fig:MgF_LR_1.8}, the Lorentzian broadening has been applied to both the spectra from the time-domain simulations and also to the stick-spectra generated from frequency-domain EOM-CCSD so that the spectra can be compared directly.

Let us begin by analyzing the linear absorption spectra at $R_\mathrm{Mg\mbox{-}F} = 1.600$ \AA. At this geometry, the energies of all states reported in Tables \ref{tab:MgF_omega_f_sigmaplus} are real-valued. The oscillations in the time-dependent dipole moment depicted in Fig.~\ref{fig:MgF_LR_1.6}(A) are stable (see the Supplementary Material for the full propagation up to $t=5000$ a.u., where the dynamics are still stable), and the resulting TD-EOM-CCSD spectrum is indistinguishable from the broadened EOM-CCSD spectrum [Fig.~\ref{fig:MgF_LR_1.6}(C)]. This agreement validates the correctness of our field-driven TD-EOM-CCSD code.

Figure \ref{fig:MgF_LR_1.8}(A) illustrates the time-dependent dipole moment for MgF at $R_\mathrm{Mg\mbox{-}F} = 1.800$ \AA\ (\emph{i.e.}, at the beginning point of the $5\;\Sigma^{+}$-- $6\;\Sigma^{+}$ complex seam in our grid of Mg--F bond distances), when excited by a field of the form of Eq.~\ref{EQN:GAUSS_SINE_FIELD}. In this case, we see that the signal is not quite stable, which is due to the presence of complex energy eigenvalues. In fact, toward the end of the TD-EOM-CCSD simulation at $t=5000$ a.u., the amplitudes of the time-dependent dipole moment oscillations is more than an order of magnitude larger than at the beginning of the simulation (see the Supplementary Material). However, this issue is not obvious if one only considers the dampened signal in Fig.~\ref{fig:MgF_LR_1.8}(B) or the linear absorption spectrum itself in Fig.~\ref{fig:MgF_LR_1.8}(C). Indeed, the absorption spectrum is, again, indistinguishable from the (real-part of the) broadened EOM-CCSD spectrum. Hence, from the spectrum alone, we have no indication that the underlying system may exhibit any unphysical properties. The only indication that something could be wrong with the spectrum is that the peak at $\sim$0.5 $\mathrm{E_h}$ is twice as tall as the corresponding peak in the EOM-CCSD stick spectrum, but this information can only be gained by comparing the time- and frequency-domain EOM-CCSD calculations, so, without performing the frequency-domain calculation, there is no \emph{a priori} way of knowing that the peak reflects the cumulative oscillator strength of two states that are degenerate (at least in the real part of the energies) or that the energies of the corresponding states are actually complex.

\subsection{Rabi cycle simulations using TD-EOM-CCSD}

\begin{figure*}[!htbp]
    \centering
    \subfigure[1.600 \AA]{\includegraphics[width=0.45\textwidth]{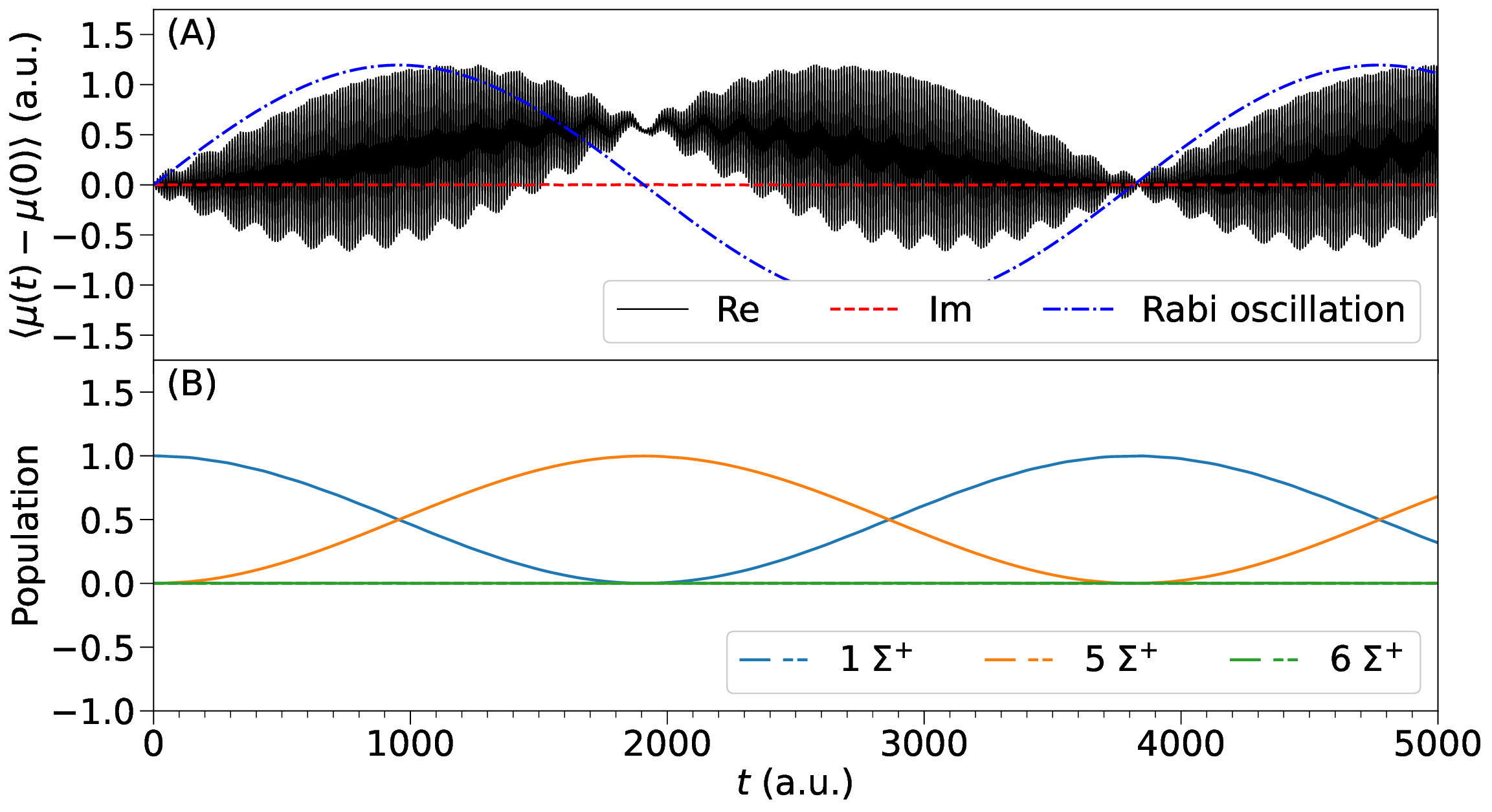}}
    \subfigure[1.692 \AA]{\includegraphics[width=0.45\textwidth]{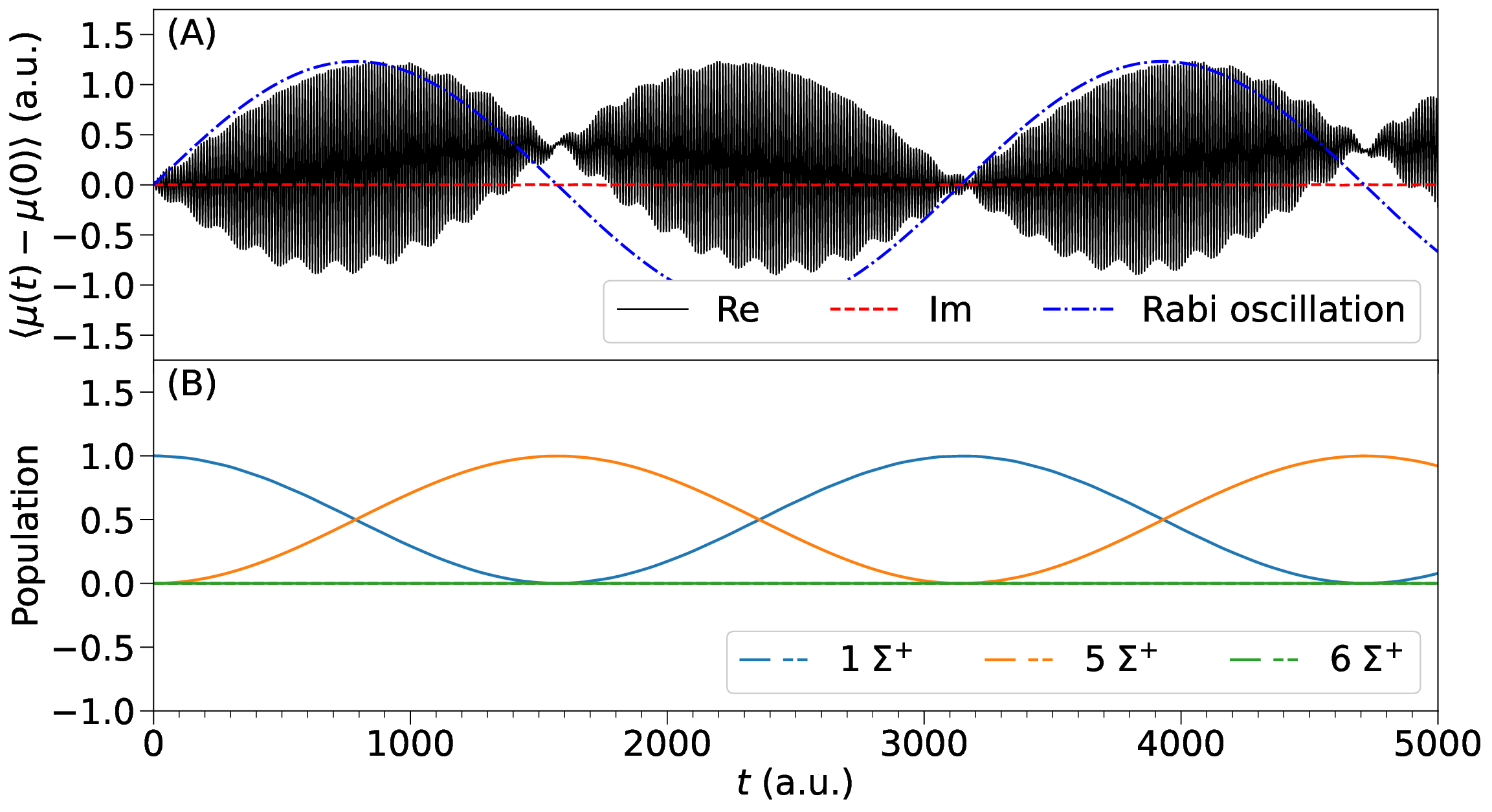}}\\[-2ex]
    \subfigure[1.790 \AA]{\includegraphics[width=0.45\textwidth]{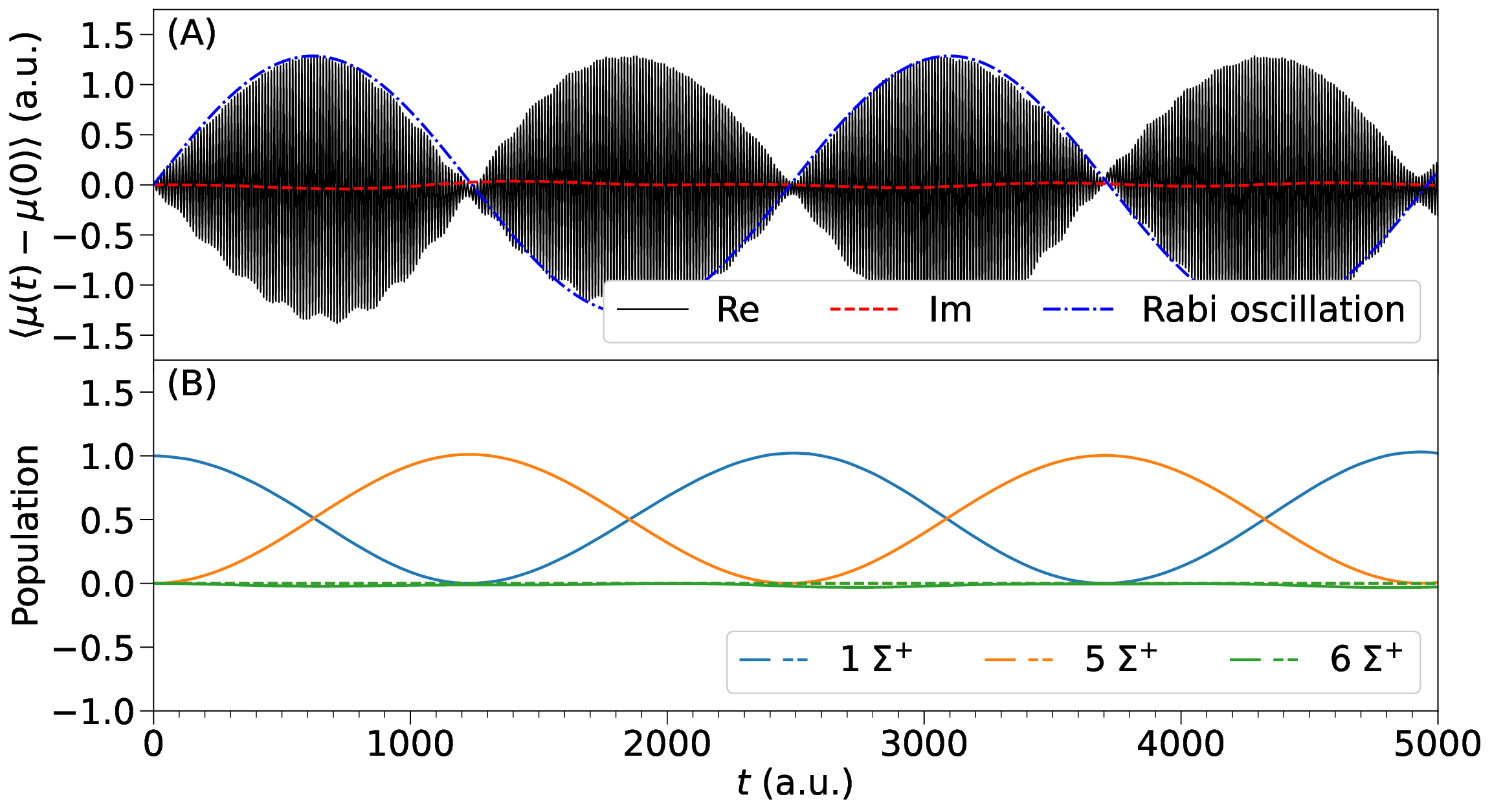}}
    \subfigure[1.800 \AA]{\includegraphics[width=0.45\textwidth]{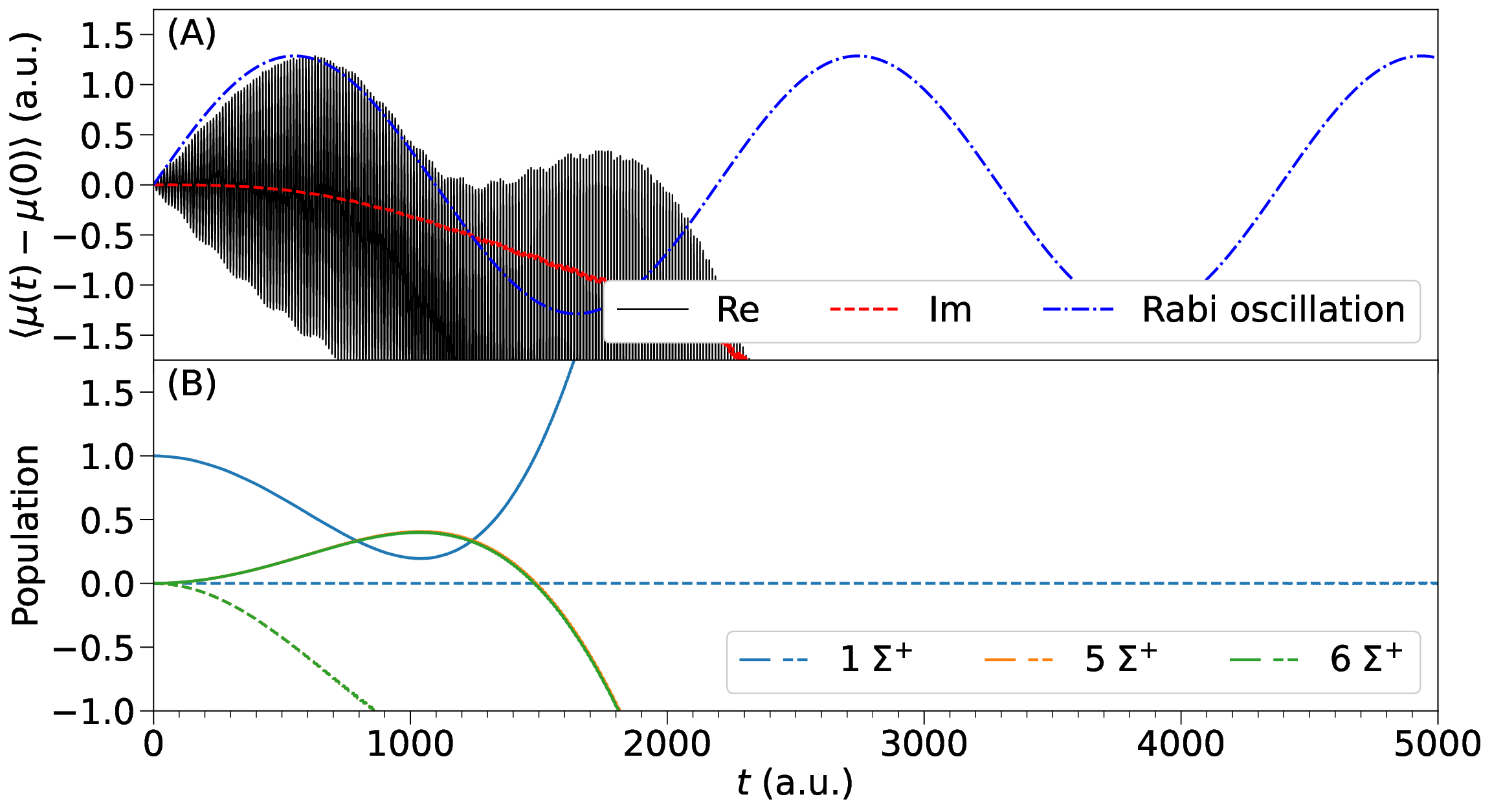}}\\[-2ex]
    \subfigure[1.820 \AA]{\includegraphics[width=0.45\textwidth]{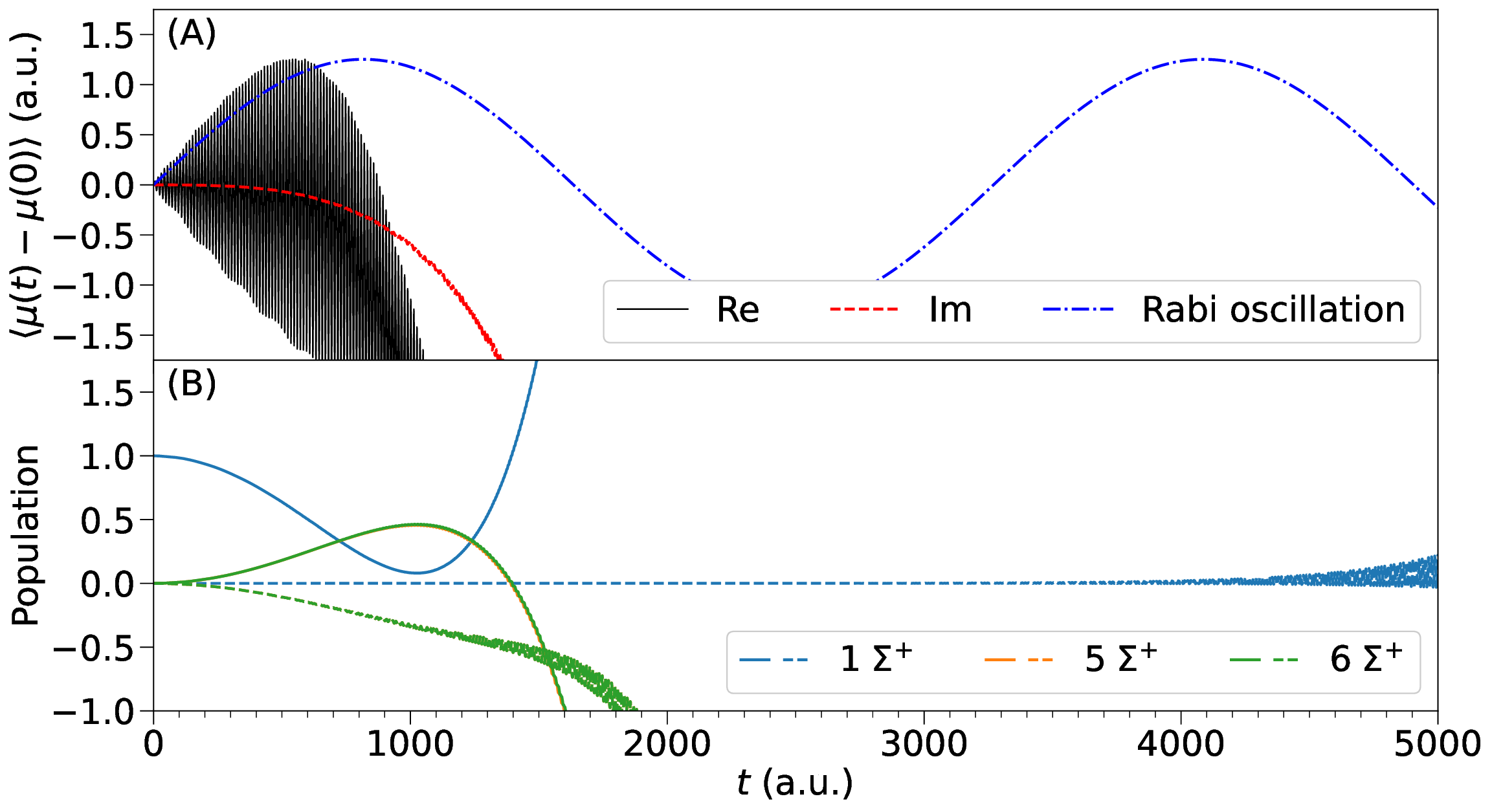}}
    \subfigure[1.830 \AA]{\includegraphics[width=0.45\textwidth]{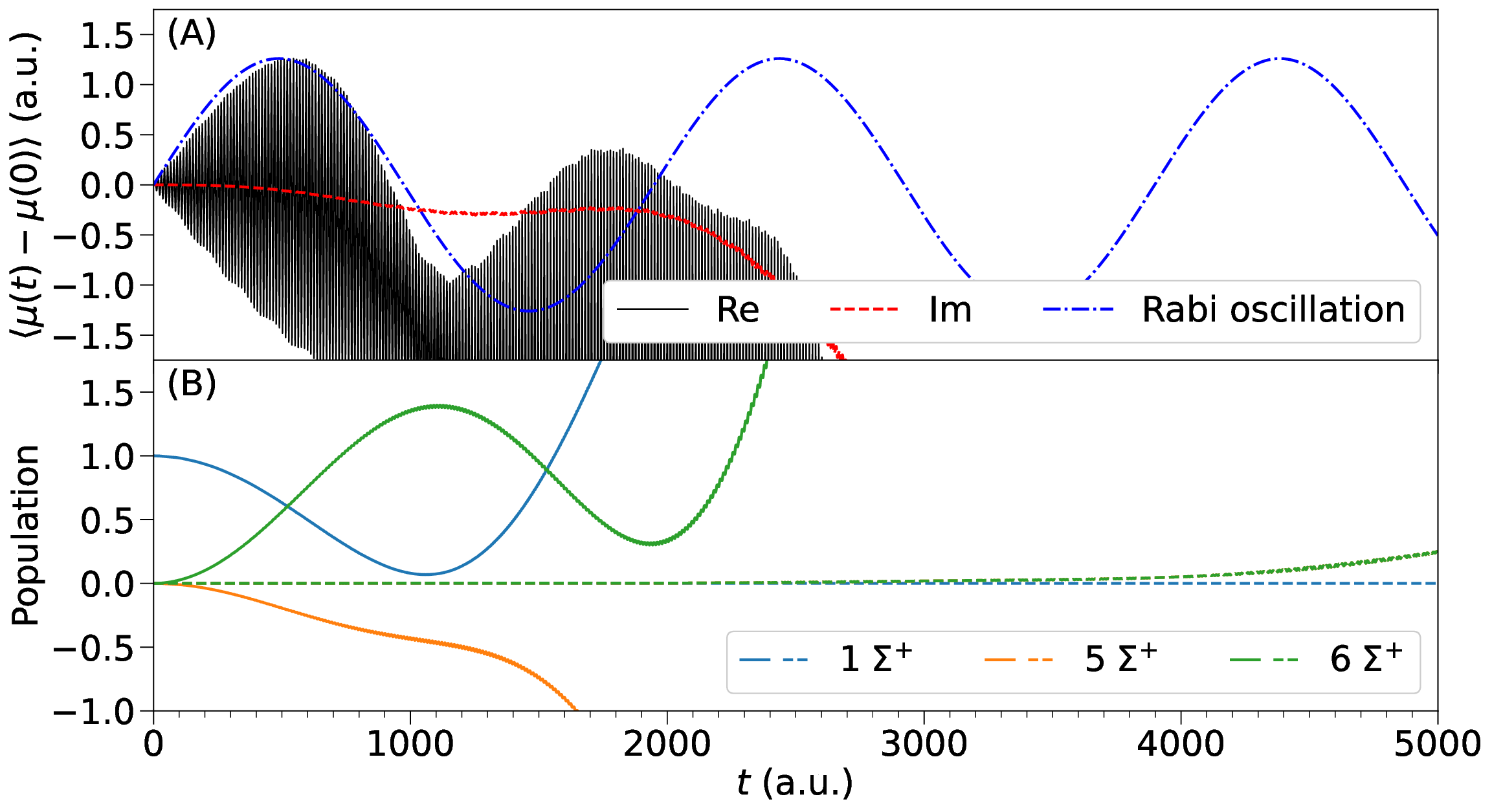}}\\[-2ex]
    \subfigure[1.900 \AA]{\includegraphics[width=0.45\textwidth]{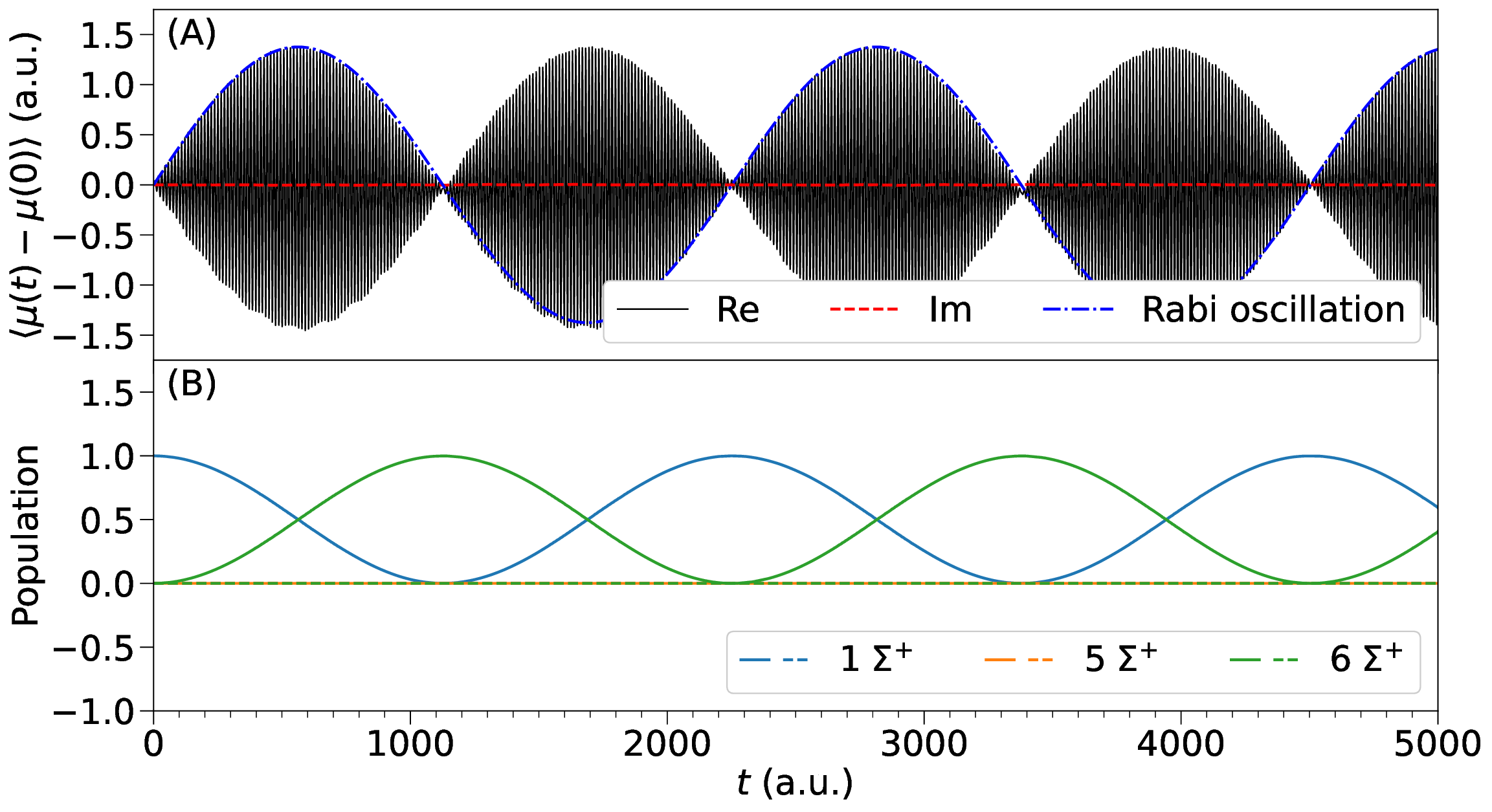}}
    \subfigure[2.000 \AA]{\includegraphics[width=0.45\textwidth]{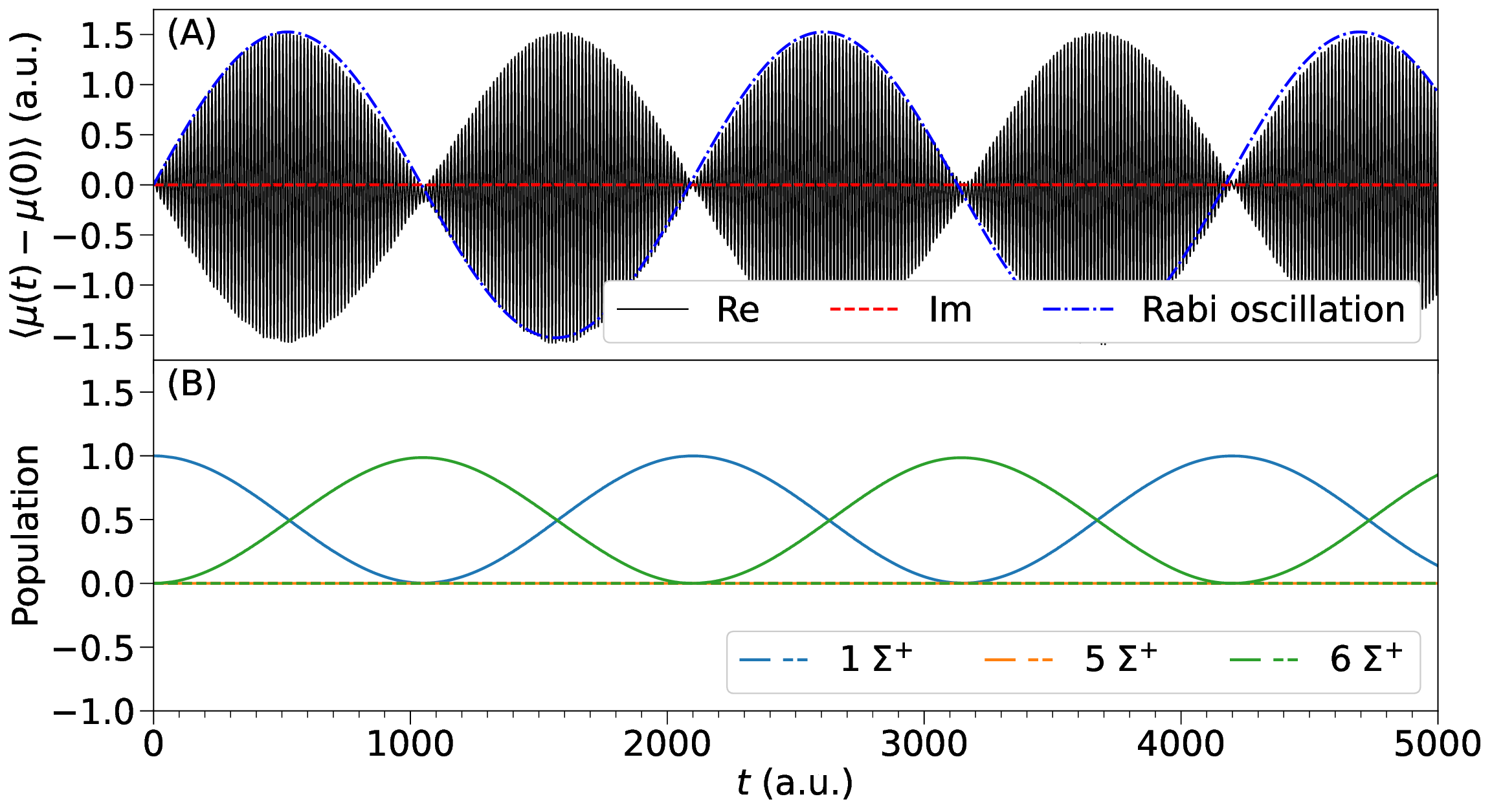}}
    \caption{The results of TD-EOM-CCSD propagations for MgF at selected Mg--F bond distances. In each panel, the upper plot (A) shows the real (black) and imaginary (red) parts of the time-dependent dipole moment expectation value along with the expected Rabi frequency (blue), while the lower plot (B) reports the real (solid line) and imaginary (dashed line) parts of the populations of the 1st, 5th, and 6th $\Sigma^{+}$ states. We used Eq.\ \ref{EQN:SINE_FIELD} with $\omega_\mathrm{ex}$ that is polarized along the $z$ axis and is resonant with the vertical excitation energies of the $5\;\Sigma^{+}$ state in panels (a)--(c), the real component of the vertical excitation energies to the $5\;\Sigma^{+}$/$6\;\Sigma^{+}$ states in panels (d)--(e), and the the vertical excitation energies of the $6\;\Sigma^{+}$ in panels (f)--(h).}
    \label{fig:MgF_rabi_8panel}
\end{figure*}

\begin{figure}[htbp]
    \centering
    \includegraphics[width=0.5\textwidth]{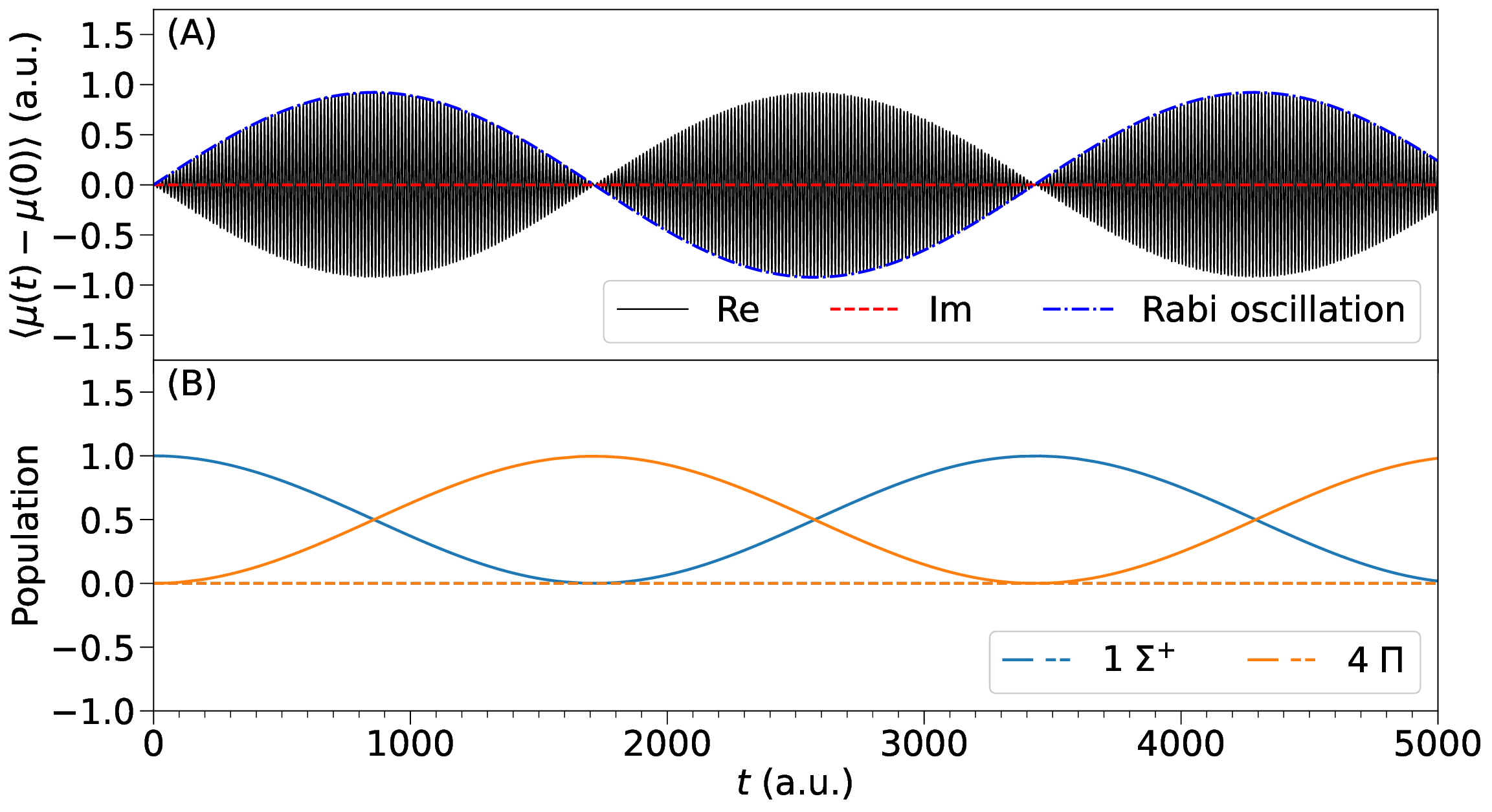}
    \caption{The field-driven TD-EOM-CCSD propagation with $\omega_\mathrm{ex}=0.4581$ a.u.\ for MgF at 1.800 \AA. (A) The real and imaginary parts of the dipole expectation value, along with the ideal Rabi oscillation computed from the corresponding transition dipole moment. (B) Real (solid) and imaginary (dashed) parts of the populations of the ground $1\;\Sigma^{+}$ and excited $4\;\Pi$ states.}
    \label{fig:MgF_1.800_Pi}
\end{figure}

\begin{figure}[htbp]
    \centering
    \includegraphics[width=0.5\textwidth]{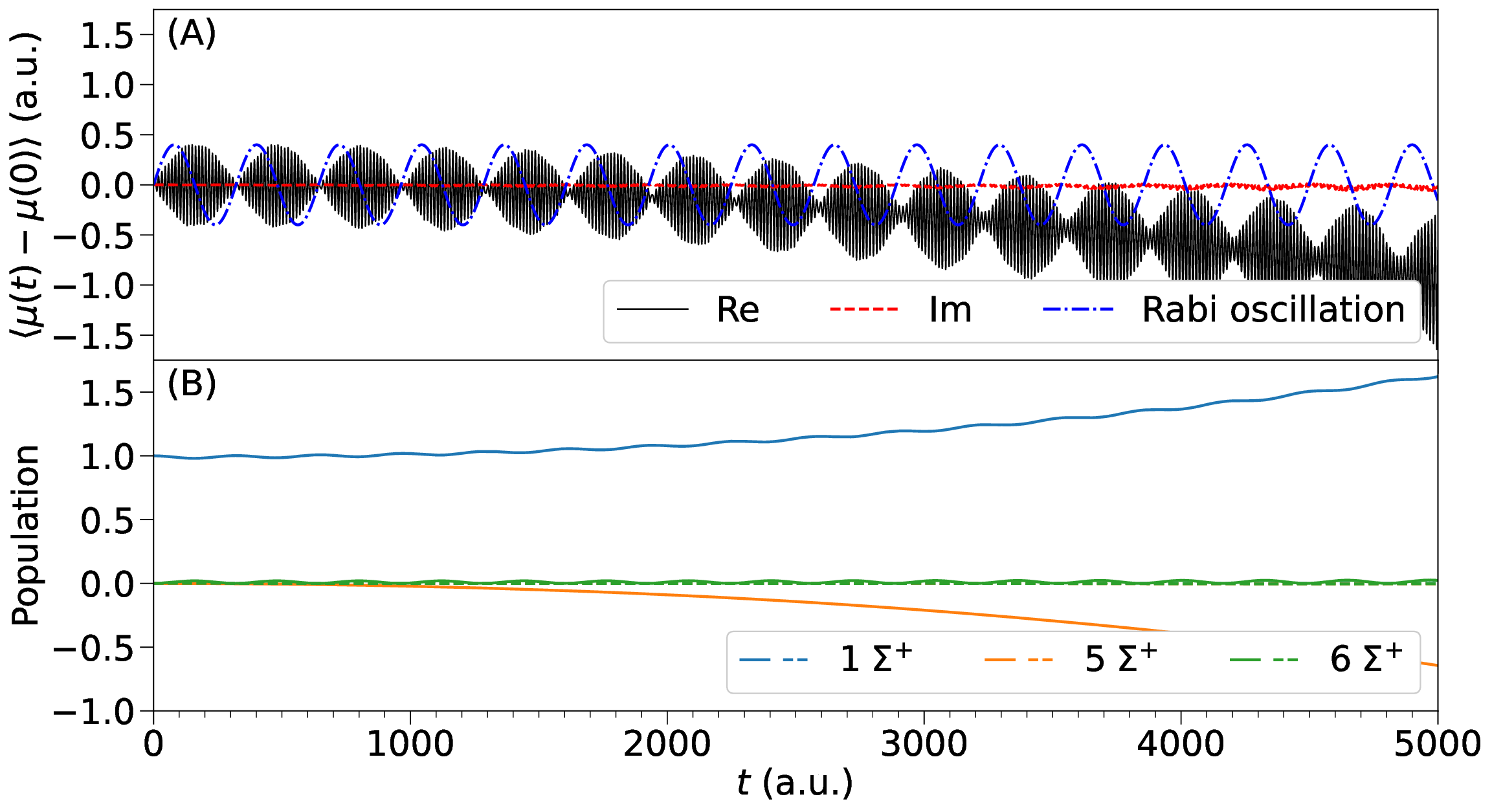}
    \caption{The field-driven TD-EOM-CCSD propagation with $\omega_\mathrm{ex}=0.4699$ a.u.\ for MgF at 1.900 \AA. (A) The real and imaginary parts of the dipole expectation value, along with the ideal Rabi oscillation computed from the corresponding transition dipole moment. (B) Real (solid) and imaginary (dashed) parts of the populations of the ground $1\;\Sigma^{+}$ and excited $5\;\Sigma^{+}$ and $6\;\Sigma^{+}$ states.}
    \label{fig:MgF_1.900_negosc}
\end{figure}

We have demonstrated that linear absorption spectra generated from TD-EOM-CC may not provide compelling evidence of the presence of complex energy eigenvalues that are clearly apparent from frequency-domain EOM-CC calculations. However, the situation is dramatically different when considering driven, resonant transitions to either states with complex-valued energies or states exhibiting negative oscillator strengths, which tend to appear in nearby regions. As is shown below, driving transitions to these states can result in unphysical phenomena, including time-dependent states composed of stationary states having populations that can be negative, exceed one, or even be complex-valued. The results of such TD-EOM-CCSD simulations on MgF with resonant, continuous-wave external field are showcased in Figs.\ \ref{fig:MgF_rabi_8panel}--\ref{fig:MgF_1.900_negosc}.

Figure \ref{fig:MgF_rabi_8panel} illustrates data from TD-EOM-CCSD simulations of Rabi oscillations at selected bond distances corresponding to those shown in Tables \ref{tab:MgF_omega_f_sigmaplus} and \ref{tab:MgF_omega_f_othersym}. In these simulations, we used external electric fields that are polarized along the $z$ axis to allow access to excited states of $\Sigma^{+}$ symmetry, with a frequency that is resonant to the $5\;\Sigma^{+}$ and $6\;\Sigma^{+}$ states at Mg--F bond distances shorter and longer than the complex seam region, respectively. In the $R_\mathrm{Mg\mbox{-}F}= 1.800$ and 1.820 \AA\ cases, where the excitation energies of these two states are complex, we used only the real part of the excitation energy as the field frequency. 

Let us begin by examining the results of a TD-EOM-CCSD simulation at an Mg--F bond distance of 1.600 \AA, where we pump to the $5\;\Sigma^{+}$ state, which are shown in panel (a) of Fig.\ \ref{fig:MgF_rabi_8panel}. First of all, at this geometry, the real part of the time-dependent dipole moment displays a well-behaved beating that matches the expected Rabi frequency, while the imaginary part stays practically zero. The Rabi frequency also matches the oscillations in the populations of the $1\;\Sigma^{+}$ and $5\;\Sigma^{+}$ states, with the population inversion happening at the mid-point of the cycle (around $t=1900$ a.u.). The population inversion also correlates with the shifted baseline of the time-dependent dipole moment after one beat. This shift reflects the fact that the dipole expectation value at this point in time corresponds to the permanent dipole moment of the $5\;\Sigma^{+}$ state, which differs from that of the ground state. We also see a smaller, more rapid beating within dipole manifold, which arises from the detuning between the field frequency and the $1\;\Sigma^{+}$-- $6\;\Sigma^{+}$ excitation energy. However, the external field is sufficiently off-resonance to prevent population transfer between the ground and $6\;\Sigma^{+}$ states, despite the fact that the oscillator strength of $6\;\Sigma^{+}$ is larger than that of $5\;\Sigma^{+}$ at 1.600 \AA\ (\emph{cf.}\ Fig.\ \ref{fig:MgF_OS} and Table \ref{tab:MgF_omega_f_sigmaplus}). 

As we stretch the Mg--F bond (and continue to pump to the $5\;\Sigma^{+}$ state), we observe a similar pattern in the dipole response at $R_\mathrm{Mg\mbox{-}F}=R_\mathrm{e}=1.692$ \AA\ [Fig.\ \ref{fig:MgF_rabi_8panel}(b)]. The Rabi frequency in this case increases slightly due to the increased oscillator strength characterizing the $5\;\Sigma^{+}$ state (to be more precise, the transition dipole moment corresponding to the $1\;\Sigma^{+}$ $\to $ $5\;\Sigma^{+}$ transition). Interestingly, as we stretch the Mg--F bond further to 1.790 \AA, we begin to see small fluctuations in the imaginary component of the dipole response signal [Fig.\ \ref{fig:MgF_rabi_8panel}(c)]. Although we still observe coherent oscillations in the populations of the $1\;\Sigma^{+}$ and $5\;\Sigma^{+}$ states, unphysical behavior in the $6\;\Sigma^{+}$ state becomes apparent. Specifically, the population of this state oscillates between zero and a small negative value, which can be attributed to the negative oscillator strength characterizing the $1\;\Sigma^{+}\rightarrow 6\;\Sigma^{+}$ excitation at this geometry. We will return to the issue of negative oscillator strength and its effect on field-driven transitions below.

We now turn to TD-EOM-CCSD simulations at Mg--F bond distances of 1.800 and 1.820 \AA, where the excitation energies of the $5\;\Sigma^{+}$ and $6\;\Sigma^{+}$ states are a complex conjugate pair. As shown in panels (d) and (e) of Fig.\ \ref{fig:MgF_rabi_8panel}, the complex excitation energy immediately interferes with the Rabi cycle. The time-dependent dipole moment at these geometries shows an exponential behavior with the imaginary component rapidly becoming negative, with a large magnitude. Interestingly, there is still a semblance of a beat at 1.800 \AA\ that matches the expected Rabi frequency. However, at 1.820 \AA, the initial beating pattern does not match the Rabi frequency calculated from the transition dipole moment. Focusing on the populations, it is worth noting that, at 1.800 \AA, the population of the ground state oscillates once and then quickly grows to be greater than one, with its imaginary component staying numerically zero. On the other hand, the imaginary components of the $5\;\Sigma^{+}$ and $6\;\Sigma^{+}$ state populations start growing from the early part of the propagation. 
The population dynamics at 1.820 \AA\ are similar, although the imaginary component of the ground-state population becomes non-negligible toward the end of the TD-EOM-CCSD propagation.

We now examine TD-EOM-CCSD simulations at $R_\mathrm{Mg\mbox{-}F}=1.830$ \AA, where we have moved away from the complex seam and the characters of the $5\;\Sigma^{+}$ and $6\;\Sigma^{+}$ states have exchanged due to the ``avoided crossing.'' While the eigenvalues of both states are already real at this geometry, the oscillator strength of the $5\;\Sigma^{+}$ state is still very negative ($-0.3275$). As shown in Fig.\ \ref{fig:MgF_rabi_8panel}(f), where we pumped to the $6\;\Sigma^{+}$ state, this property, combined with the fact that the $5\;\Sigma^{+}$ and $6\;\Sigma^{+}$ states are energetically close (separated by roughly 2 mE$_{\rm h}$), results in time-dependent dipole moments and state populations that resemble those obtained within the complex seam. On the other hand, as we move farther from the geometries with complex eigenvalues, we recover the coherent Rabi cycle structures at 1.900 and 2.000 \AA~[panels (g) and (h), respectively, of Fig.\ \ref{fig:MgF_rabi_8panel}].

In order to verify that our observations with the complex eigenvalues are not a result of instability in our code, we performed another TD-EOM-CCSD propagation at $R_\mathrm{Mg\mbox{-}F}=1.800$ \AA, this time with the field polarized along the $x$ axis and resonant with the ground state $\to $ $4\;\Pi$ state transition. The results of this simulation are shown in Fig.\ \ref{fig:MgF_1.800_Pi}. Owing to the large separation between the $4\;\Pi$ state and the other $\Pi$ states, the Rabi oscillations shown in Fig.\ \ref{fig:MgF_1.800_Pi} are characterized by a well-defined beating and perfect agreement with the expected Rabi frequency. These results suggest that the poor behavior shown in Fig.\ \ref{fig:MgF_rabi_8panel}(d) results from the complex energy eigenvalues, as opposed to some numerical instability in the code.

Our last examination pertains to pumping to a state with negative oscillator strength. To that end, we have chosen to pump the system from the ground state to the $5\;\Sigma^{+}$ state at an Mg--F separation of 1.900 \AA; the results of this simulation are shown in Fig.\ \ref{fig:MgF_1.900_negosc}. The population of the target state becomes increasingly negative as the simulation progresses, which is consistent with what was observed in Fig.\ \ref{fig:MgF_rabi_8panel}(c) and (f). In order to compensate for the negative population of the excited state, the ground-state population (initially at 1), increases as the simulation proceeds. It is worth noting that, at this geometry, the $6\;\Sigma^{+}$ state has a much larger oscillator strength than the target $5\;\Sigma^{+}$, which results in a beating pattern that has a significant contribution from the detuned frequency.

{\color{black}
\subsection{EOM-CCSD with a symmetrized $\bar{H}$}

{\color{black}It is worth noting that several strategies (see, \emph{e.g.}, Ref.\ \onlinecite{Koch17_4801}) have been put forth to remedy defects in the similarity-transformed Hamiltonian when it is represented in a truncated many-electron basis. However, these approaches require {\em a priori} knowledge that $\bar{H}$ is defective, which renders them less helpful in the context of TD-EOM-CC calculations performed in a basis of Slater determinants, as opposed to the basis that diagonalizes $\bar{H}$. On the other hand, a straightforward way to guarantee real eigenvalues (and non-negative oscillator strengths), without the need to diagonalize $\bar{H}$, would be to simulate electron dynamics using a symmetrized similarity-transformed Hamiltonian matrix, {\em i.e.}, $\tilde{\bar{H}} = \frac{1}{2} \left( \bar{H} + \bar{H}^T \right)$. Indeed, this is precisely the strategy used in some earlier TD-EOM-CC\cite{Schlegel11_4678,Head-gordon12_909} and more recent TD-CC\cite{Schoyen21_388} studies. As in those studies, we consider light-matter interactions mediated by a symmetrized similarity-transformed dipole matrix, $\tilde{\bar{\mu}} = \frac{1}{2} \left( \bar{\mu} + \bar{\mu}^T \right)$.}

\begin{figure}[!htbp]
    \centering
    \includegraphics[width=0.5\textwidth]{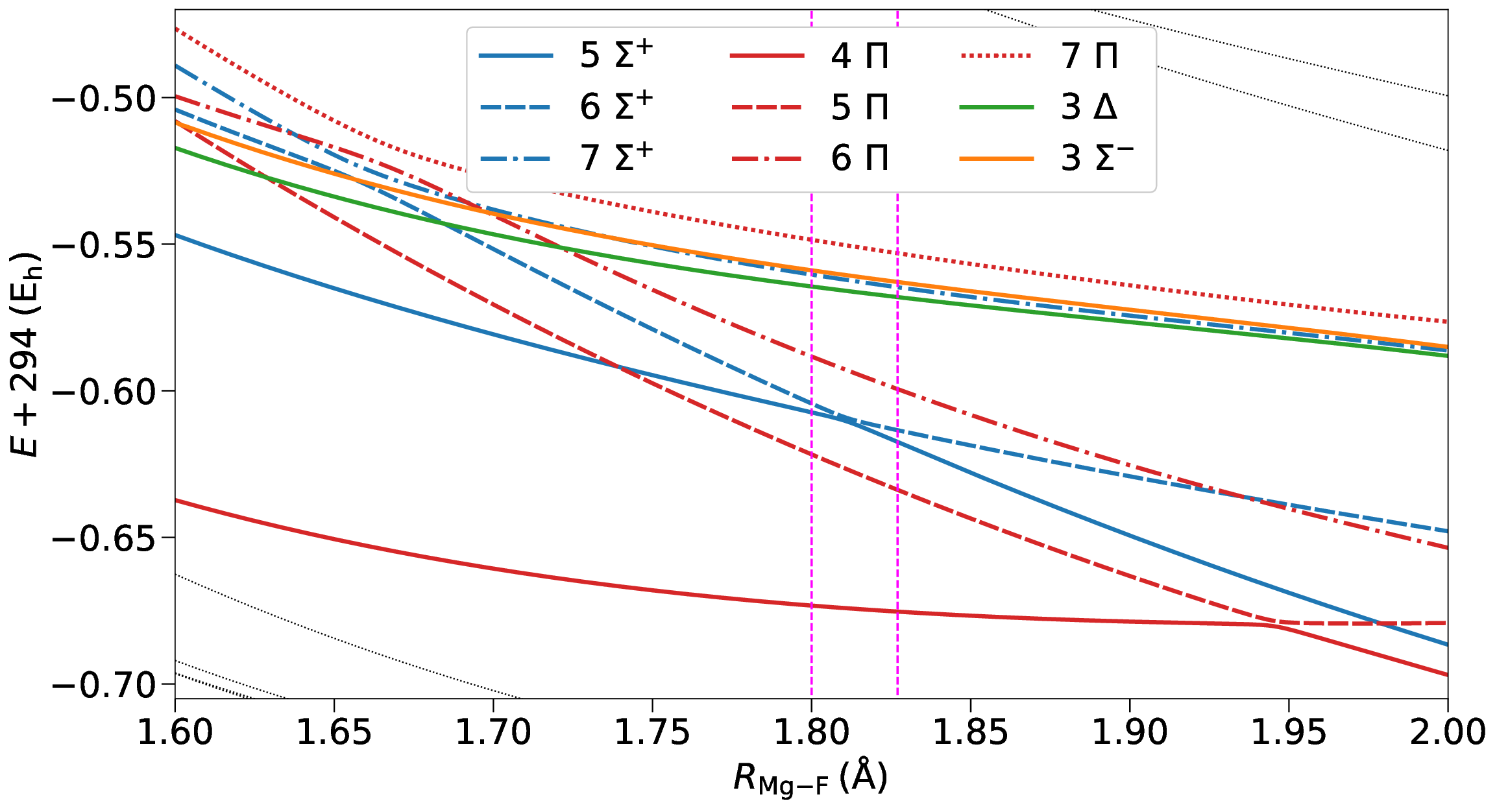}
    \caption{\color{black} Potential energy curves of selected excited states of MgF computed at the EOM-CCSD/STO-3G level of theory using the symmetric CCSD $\tilde{\bar{H}}$. The states are labeled according to increasing energy at $R_\mathrm{Mg\mbox{-}F} = \tilde{R}_\mathrm{e} = 1.694$ \AA. Note the disappearance of the EOM-CCSD complex seam between the $5\;\Sigma^{+}$ and $6\;\Sigma^{+}$ states (blue solid and dashed lines, respectively) in the [1.800,1.827] \AA\ region, which is demarcated with vertical magenta lines (\emph{cf.}\ Fig.\ \ref{fig:MgF_PEC}. Other states that are energetically far from the relevant states are marked with black dotted lines.}
    \label{fig:MgF_PEC_hermitian}
\end{figure}

\begin{figure}[!htbp]
    \centering
    \includegraphics[width=0.5\textwidth]{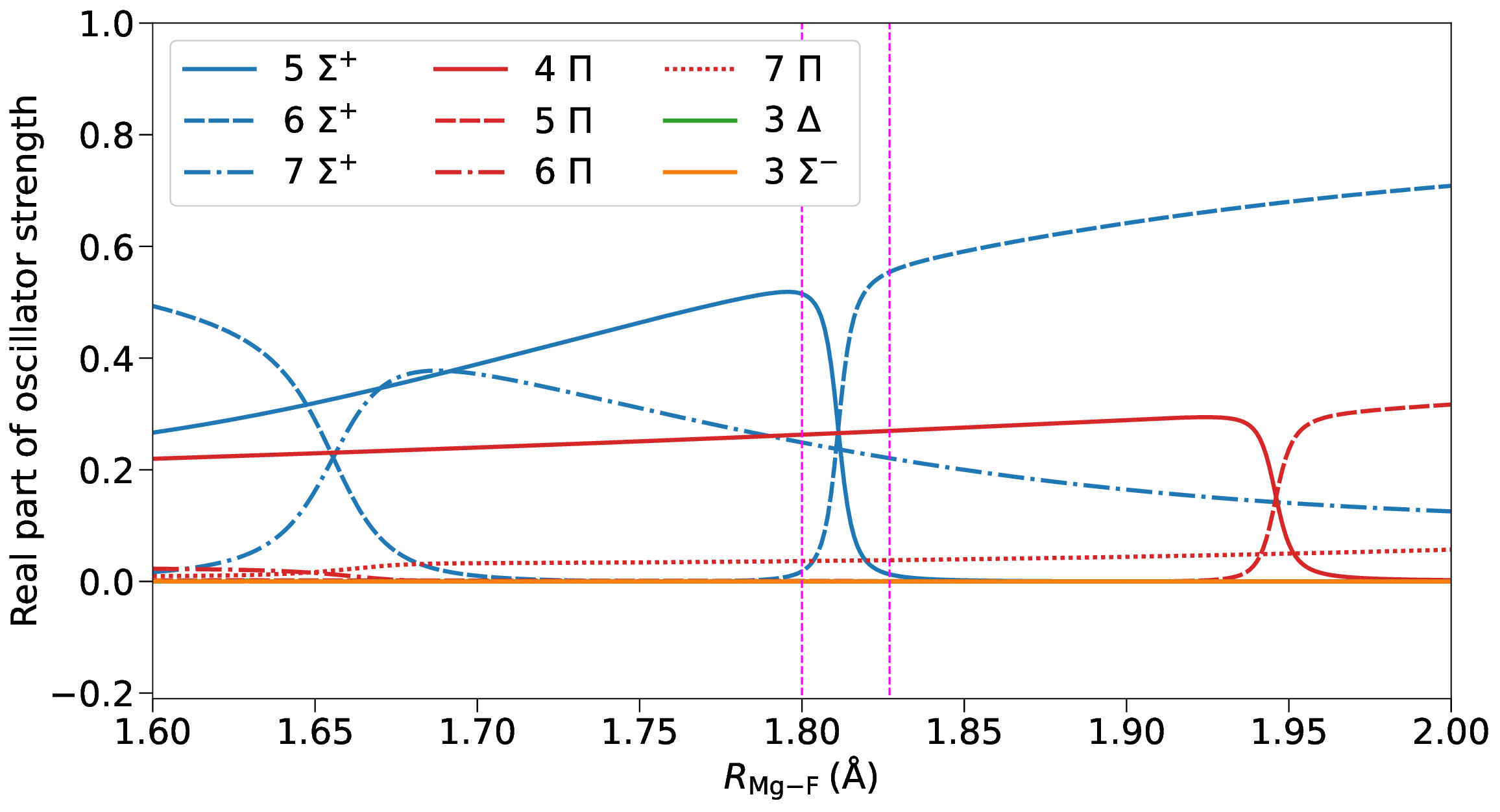}
    \caption{\color{black} Oscillator strength curves of selected excited states of MgF computed at the the EOM-CCSD/STO-3G level of theory using symmetric CCSD $\tilde{\bar{H}}$ and $\tilde{\bar{\mu}}$. The states are labeled according to increasing energy at $R_\mathrm{Mg\mbox{-}F} = \tilde{R}_\mathrm{e} = 1.694$ \AA. Note the disappearance of the EOM-CCSD complex seam between the $5\;\Sigma^{+}$ and $6\;\Sigma^{+}$ states (blue solid and dashed lines, respectively) in the [1.800,1.827] \AA\ region, which is demarcated with vertical magenta lines (\emph{cf.}\ Fig.\ \ref{fig:MgF_OS} ). Other states that are energetically far from the complex seam are marked with black dotted lines.}
    \label{fig:MgF_OS_hermitian}
\end{figure}

{\color{black}Figure \ref{fig:MgF_PEC_hermitian} demonstrates that the eigenvalues of $\tilde{\bar{H}}$ are real-valued and that the avoided crossing between the $5\:\Sigma^{+}$ and $6\:\Sigma^{+}$ states in the [1.800,1.827] \AA\ region is well described. Aside from this feature, the energies obtained by diagonalizing $\bar{H}$ and $\tilde{\bar{H}}$ are quite similar.  If we focus on the 5th, 6th, and 7th $\Sigma^{+}$ states at 1.692 \AA, the differences between the unmodified and symmetrized EOM-CCSD total energies are only 0.010, 0.635, and 0.371 mE$_{\rm h}$, respectively. In addition, the ground-state equilibrium bond distances are practically identical in the unmodified and symmetrized cases (1.692 and 1.694 \AA, respectively).

As for oscillator strenghts, those provided in Fig.~\ref{fig:MgF_OS_hermitian} appear similar to those depicted in Fig.~\ref{fig:MgF_OS}. We can also see the clear transfer of character between the oscillator strengths for the $5\:\Sigma^{+}$ and $6\:\Sigma^{+}$ states around the avoided crossing in Fig.~\ref{fig:MgF_OS_hermitian}, which was not observed in Fig.~\ref{fig:MgF_OS}. Moreover}, the expected Rabi frequencies associated with the $1\;\Sigma^{+}\rightarrow 5\;\Sigma^{+}$ transition at the Mg--F separation of 1.692 \AA\ derived from $\bar{H}$ and $\bar{\mu}$ versus $\tilde{\bar{H}}$ and $\tilde{\bar{\mu}}$ are  $1.999\times10^{-3}$ and $1.994\times10^{-3}$ a.u., respectively (assuming $\vert\va{\varepsilon}_\mathrm{max}\vert = 2.0\times10^{-3}$ a.u.). This observation suggests that dynamics obtained from TD-EOM-CCSD simulations based on $\bar{H}$ and $\tilde{\bar{H}}$, at geometries far from the complex seam, should be quite similar. 

Despite the promising properties of $\tilde{\bar{H}}$, several key details suggest that this strategy may not be advisable in general. First, we note that the \emph{ground-state} energies of $\tilde{\bar{H}}$ do differ from $E_0$ by a non-trivial amount (on the order of $\sim$0.01 E$_{\rm h}$ or 0.27 eV for MgF/STO-3G). Related to these changes is the fact that neither the ground-state right- nor left-hand CC wave functions are stationary states of $\tilde{\bar{H}}$. Consequently, physically meaningful dynamics out of the ground state requires that one recompute the ground-state of $\tilde{\bar{H}}$ prior to any TD-EOM-CC simulations. Second, the more notable deficiency of this approach is that a symmetrized EOM-CC approach does not converge to the full CI limit. We demonstrate this failure numerically for a case for which EOM-CCSD is equivalent to the full CI: molecular hydrogen described by the cc-pVDZ basis set. Table \ref{tab:H2_spectra} shows the lowest 10 eigenvalues of $\bar{H}$ and $\tilde{\bar{H}}$ constructed from CCSD cluster amplitudes for H$_2$, with an H--H separation of 1.0 \AA. As the data show, none of the eigenvalues are preserved, and the errors in $\tilde{\bar{H}}$ are not uniform in sign or magnitude. Even worse, the error in the ground-state energy is roughly 0.01 E$_{\rm h}$.
}

\begin{table}[!htbp]
    \caption{\label{tab:H2_spectra} \color{black} Several lowest eigenvalues (in $\mathrm{E_h}$) of H$_2$ at an H--H interatomic distance of 1.0 \AA, obtained by diagonalizing the conventional ($\bar{H}$) and symmetrized ($\tilde{\bar{H}}$) CCSD similarity-transformed Hamiltonian matrices, along with the difference between each pair of eigenvalues.}
    
    \centering
    \begin{tabular}{c @{\extracolsep{8pt}}c @{\extracolsep{8pt}}c @{\extracolsep{8pt}}c}
        \hline\hline
        State & $E[\bar{H}]$ & $E[\tilde{\bar{H}}]$ & $E[\tilde{\bar{H}}]-E[\bar{H}]$ \\
        \hline
        0 & $-1.140073$ & $-1.149699$ & $-0.009626$ \\
        1 & $-0.876810$ & $-0.876878$ & $-0.000068$ \\
        2 & $-0.702810$ & $-0.702872$ & $-0.000063$ \\
        3 & $-0.488230$ & $-0.488263$ & $-0.000034$ \\
        4 & $-0.363672$ & $-0.362644$ &   0.001028  \\
        5 & $-0.335016$ & $-0.333321$ &   0.001695  \\
        6 & $-0.281840$ & $-0.281901$ & $-0.000061$ \\
        7 & $-0.197986$ & $-0.198026$ & $-0.000040$ \\
        8 & $-0.035498$ & $-0.035465$ &   0.000033  \\
        9 &   0.114365  &   0.114299  & $-0.000065$ \\
        \hline\hline
    \end{tabular}
\end{table}

\section{Conclusions}
\label{SEC:CONCLUSIONS}

We have observed the occurrence of complex energy eigenvalues in truncated CC/EOM-CC calculations, which result from the failure of EOM-CCSD to describe an avoided crossing (between the $5\;\Sigma^{+}$ and $6\;\Sigma^{+}$ states in the MgF/STO-3G system). As the Mg--F bond distance approaches the critical region where the energy eigenvalues become complex, the oscillator strengths of the $5\;\Sigma^{+}$ and $6\;\Sigma^{+}$ states diverge, before tilting into the complex plane, and the character transfer between the two states occurs on the imaginary axis. It is also worth emphasizing that, although the energies of the two states seem to transition ``smoothly'' from the real to the complex plane, other properties (here, exemplified by oscillator strengths) diverge as the Mg--F bond distance approaches the complex seam. This behavior indicates that the quality of the EOM-CCSD wave functions deteriorates significantly in this region.

A comparison between the linear absorption spectra obtained from frequency- and time-domain EOM-CCSD calculations at a geometry that does not exhibit complex eigenvalues shows an excellent agreement in peak location and relative peak heights. Surprisingly, the same observation can be made even at the Mg--F bond distance with complex eigenvalues, which indicates that the linear-response spectrum obtained using TD-EOM-CCSD is not significantly affected by the presence of complex energy eigenvalues. The situation is dramatically different, though, for  simulations outside of the linear response regime.

For driven excitations in the MgF/STO-3G system, we observed the well-resolved beating of the time-dependent dipole moment that we expect for the Rabi cycle, provided that (i) the Mg--F bond length is sufficiently far from the complex seam or (ii) the external field pumps to a state with a different symmetry than that of the excited states having complex eigenvalues. Resonant driven excitations to a state with complex eigenvalues introduces an exponential behavior to the time-dependent dipole expectation value and the relevant $\bar{H}$ eigenstate populations. Interestingly, the total norm of the time-dependent state in such situations is conserved, but the populations of the stationary states that make up the time-dependent state can acquire unphysical values (greater than one, negative, or complex-valued).

Pumping to a state with a real-valued energy, but a negative oscillator strength, results in a ``forced'' population transfer from the target state to the ground state. If the simulation starts with the system in the ground state, the population of the target state becomes negative, which is clearly unphysical. This phenomenon could also pose a problem when simulating transient or excited-state absorption spectra, because states with negative oscillator strengths could lead to artificial and unphysical ``emission'' to {\color{black} the} ground state. 

In summary, these observations highlight the value of a rigorous examination of the potential surfaces obtained from truncated EOM-CC calculations before performing electron dynamics simulations. Moreover, we caution against the use of linear response TD-EOM-CC calculations for this purpose, because unphysical behaviors arising due to complex eigenvalues or negative oscillator strength values may not be immediately apparent from spectra obtained from such simulations without making direct comparisons to the results of frequency-domain computations. As already alluded to in Sec.\ \ref{SEC:INTRODUCTION}, we can anticipate these undesirable behaviors to be mitigated as we employ higher-order CC/EOM-CC approaches; future work should confirm that this is indeed the case. Alternatively, because the unphysical behaviors we observe stem from the non-Hermiticity of the similarity-transformed Hamiltonian (expanded in an incomplete many-electron basis), it could be interesting to explore the utility of Hermitian alternatives to TD-EOM-CC based on unitary CC theory.\cite{Kutzelnigg77_129,Bartlett89_133,Bartlett06_3393}
{\color{black} Along these lines, one may be tempted to simply symmetrize the similarity-transformed Hamiltonian constructed in a truncated CC calculation prior to performing TD-EOM-CC propagations, but doing so introduces problems, the most severe of which is the inability of the approach to converge to the full CI limit. Instead, one could consider other strategies aimed at reducing or eliminating the non-Hermiticity of $\bar{H}$ obtained in a truncated CC calculation, while still properly defining a ground-state CC wave function, such as those outlined in Ref.\ \onlinecite{Koch17_4801}.
}

\vspace{0.5cm}

{\bf Supplementary Material} The PECs of selected excited states of MgF computed using EOM-CCSD/STO-3G with restricted open-shell HF reference, stability analysis of the code with time steps $\Delta t=0.002$, 0.01, and 0.02 a.u., comparisons between Rabi oscillations using field strength $\vert\va{\varepsilon}_\mathrm{max}\vert = 5\times 10^{-4}$, $2\times 10^{-3}$, and $5\times 10^{-3}$ a.u., and linear-response signals with total propagation times of 5000 a.u.

\vspace{0.5cm}

\begin{acknowledgments}This material is based upon work supported by the U.S. Department of Energy, Office of Science, Office of Advanced Scientific Computing Research and Office of Basic Energy Sciences, Scientific Discovery through the Advanced Computing (SciDAC) program under Award No. DE-SC0022263. This project used resources of the National Energy Research Scientific Computing Center, a DOE Office of Science User Facility supported by the Office of Science of the U.S. Department of Energy under Contract No. DE-AC02-05CH11231 using NERSC award ERCAP-0024336.
\\ 
\end{acknowledgments}

\noindent {\bf DATA AVAILABILITY}\\

The data that support the findings of this study are available from the corresponding author upon reasonable request.

\bibliography{Journal_Short_Name,main,deprince,cc}

\end{document}


\author{Stephen H. Yuwono}
\affiliation{
             Department of Chemistry and Biochemistry,
             Florida State University,
             Tallahassee, FL 32306-4390}          

\author{Brandon C. Cooper}
\affiliation{
             Department of Chemistry and Biochemistry,
             Florida State University,
             Tallahassee, FL 32306-4390}       

\author{Tianyuan Zhang}
\affiliation{Department of Chemistry, University of Washington, Seattle, WA 98195}

\author{Xiaosong Li}
\affiliation{Department of Chemistry, University of Washington, Seattle, WA 98195}
             
\author{A. Eugene DePrince III}
\email{adeprince@fsu.edu}
\affiliation{
             Department of Chemistry and Biochemistry,
             Florida State University,
             Tallahassee, FL 32306-4390}

\title{Supplementary Material for: Time-Dependent Equation-of-Motion Coupled-Cluster Simulations with a Defective Hamiltonian}

\maketitle

\pagebreak


\begin{figure}[!htbp]
    \centering
    \includegraphics[width=0.75\textwidth]{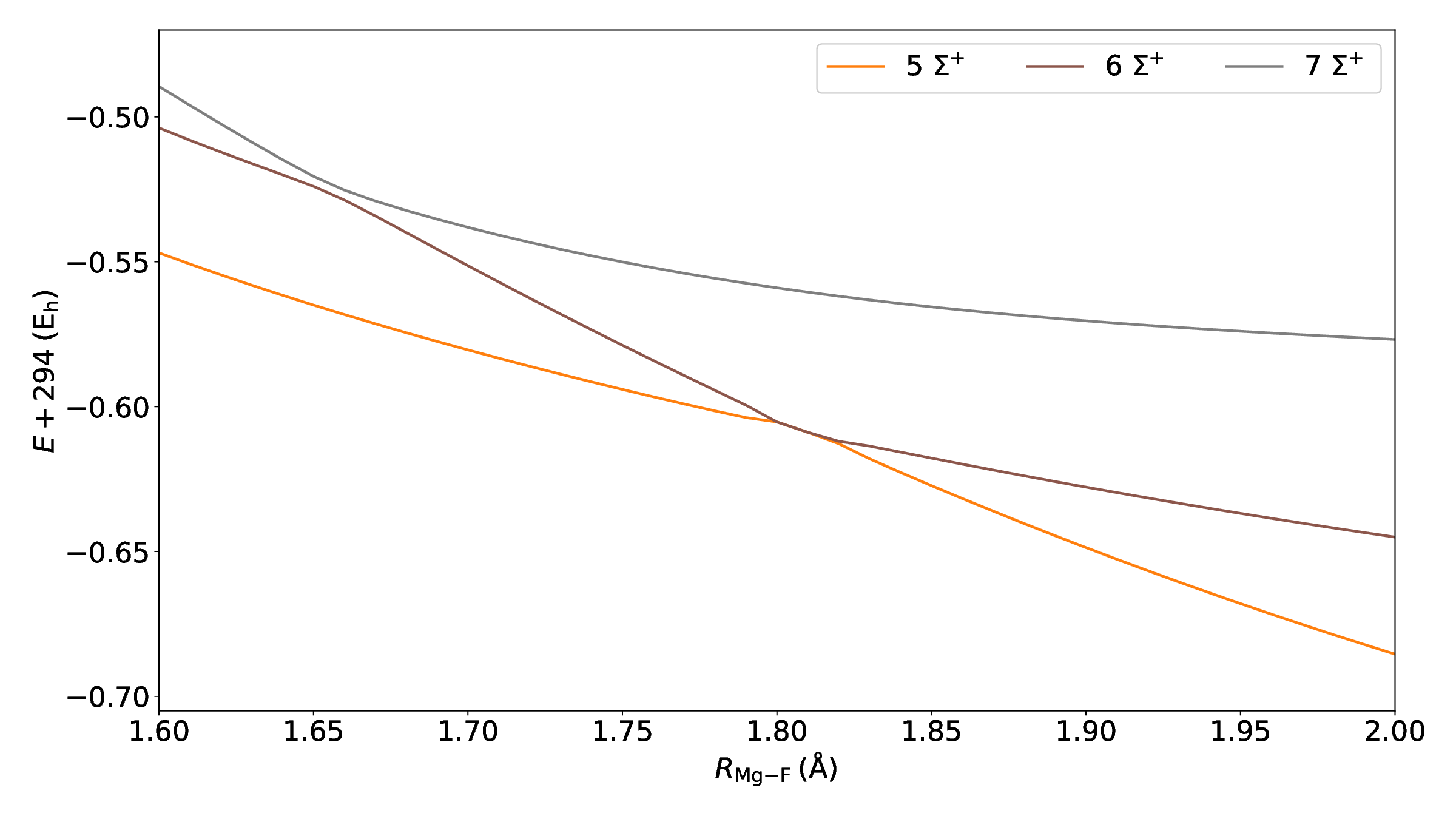}
    \caption{Potential energy curves of the $5\;\Sigma^{+}$, $6\;\Sigma^{+}$, and $7\;\Sigma^{+}$ states of MgF resulting from the EOM-CCSD/STO-3G computations with restricted open-shell Hartree--Fock (ROHF) reference. Note the occurrence of complex eigenvalues resulting from the failure of ROHF-based EOM-CCSD to properly describe avoided crossing between the $5\;\Sigma^{+}$ and $6\;\Sigma^{+}$ states (orange and brown, respectively) in the [1.80,1.82] \AA\ region.}
    \label{fig:MgF_PEC_ROHF}
\end{figure}

\begin{figure}[!htbp]
    \centering
    \includegraphics[width=0.75\textwidth]{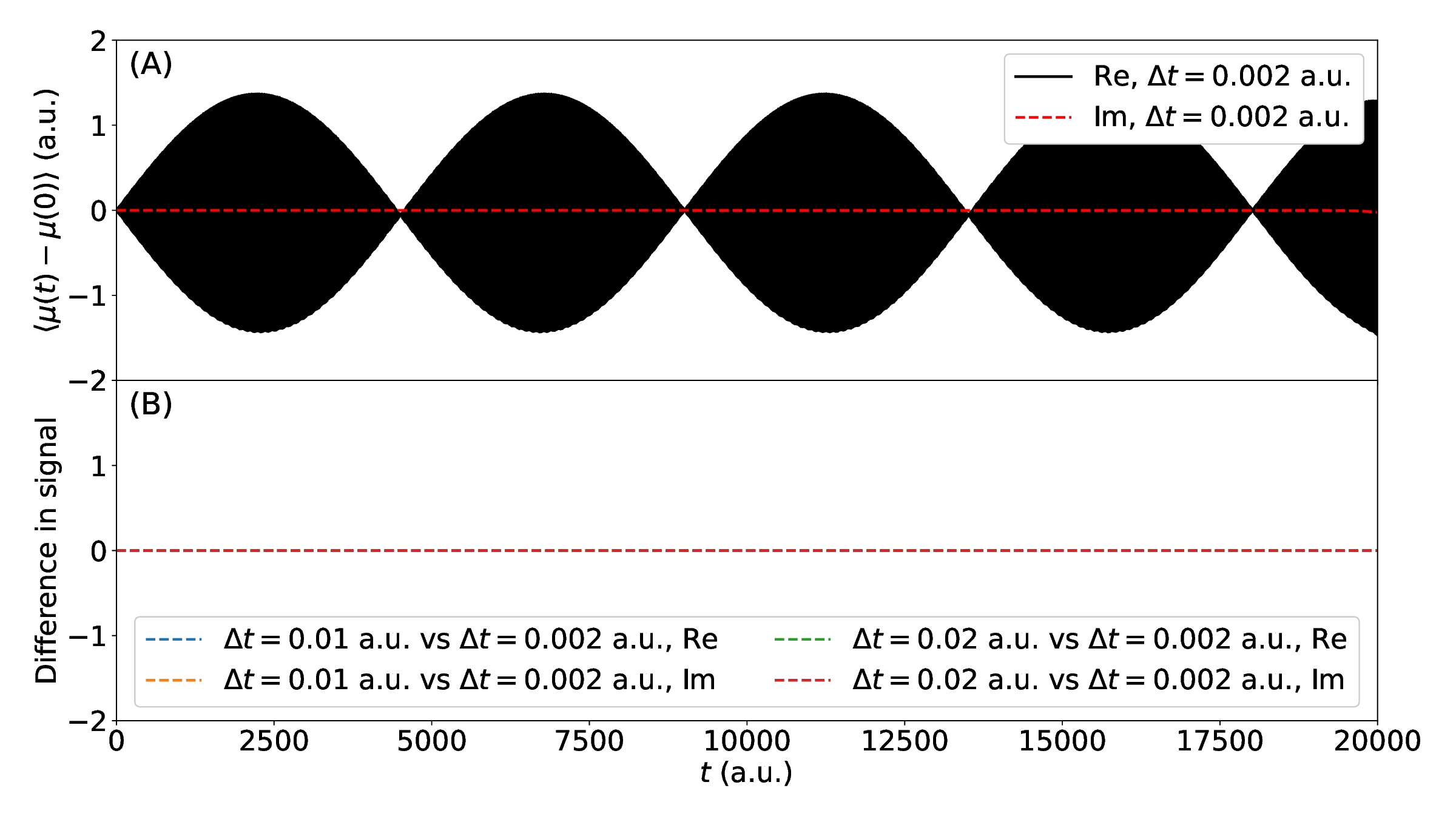}
    \caption{Stability test of our TD-EOM-CCSD propagations with respect to time step size. (A) Rabi oscillation obtained by pumping to the $6\;\Sigma^{+}$ state of MgF at the inter-atomic distance of 1.900 \AA, using a field strength of 0.0005 a.u.\ and time step of 0.002 a.u. (B) Difference in the dipole response signal obtained with larger time steps (0.01 and 0.02 a.u.).}
    \label{fig:MgF_dt_test}
\end{figure}

\begin{figure}[!htbp]
    \centering
    \includegraphics[width=0.75\textwidth]{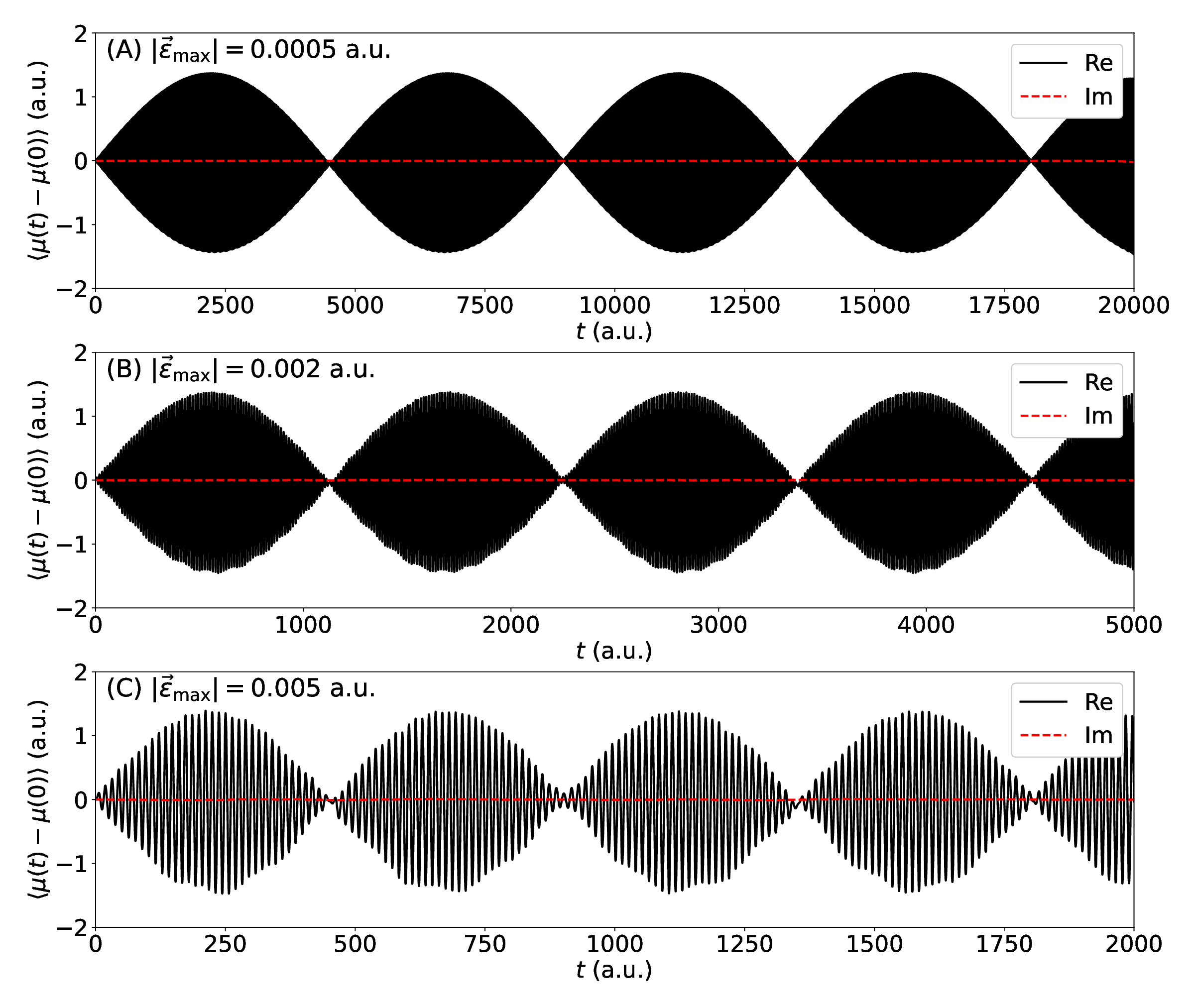}
    \caption{Rabi oscillations for MgF at 1.900 \AA\ obtained at different field strength, using time step of 0.01 a.u. The period of the beats is inversely proportional to the field strength, as expected from the definition of Rabi frequency (see main text).}
    \label{fig:MgF_field_test}
\end{figure}

\begin{figure}[!htbp]
    \centering
    \includegraphics[width=0.75\textwidth]{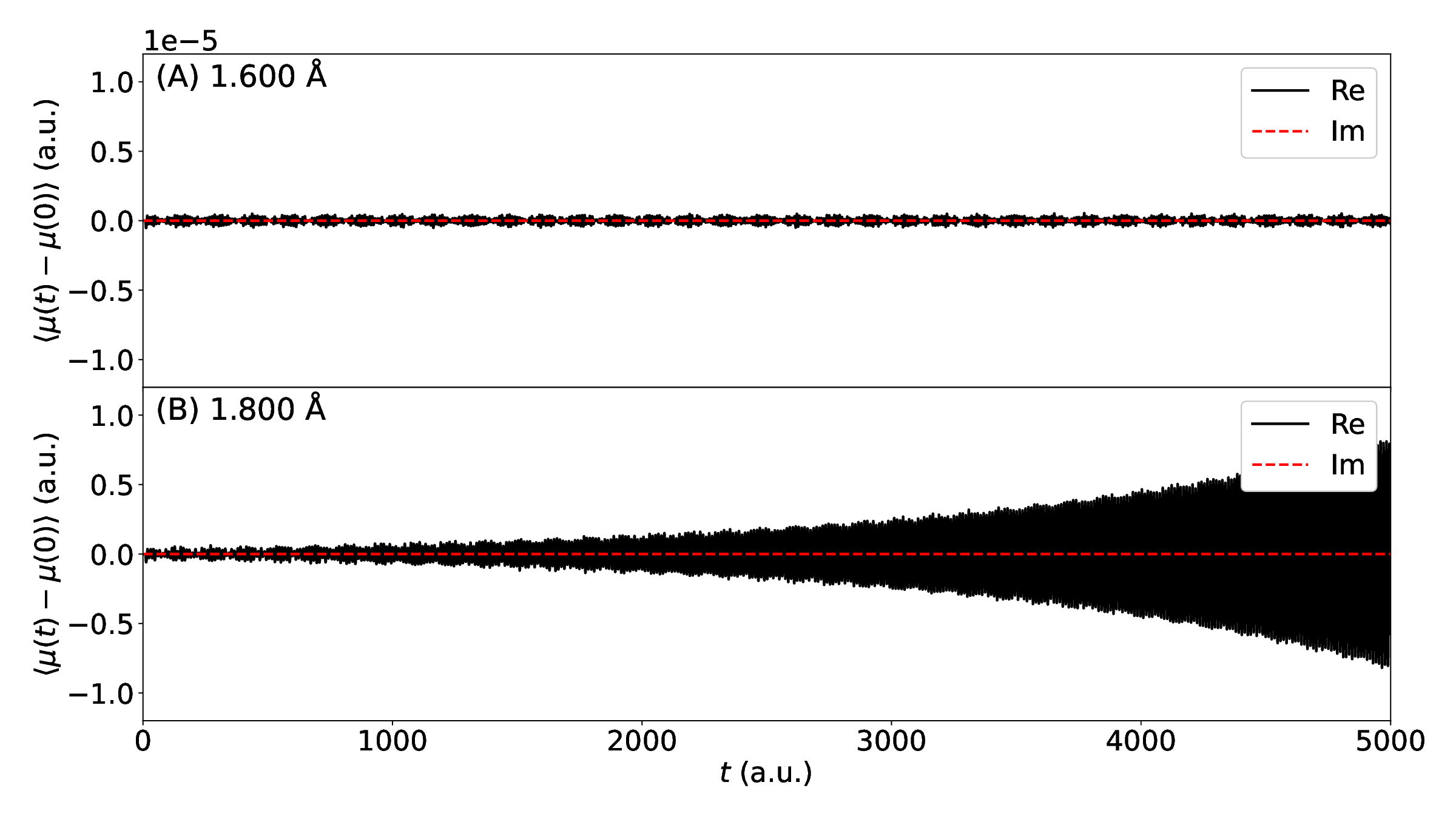}
    \caption{The full dipole response signal (up to $t=5000$ a.u.) used to obtain the linear absorption spectra of MgF at (A) 1.600 and (B) 1.800 \AA\ reported in the main text. At 1.800 \AA, the signal showed an exponential behavior due to the presence of complex eigenvalues.}
    \label{fig:MgF_LR_full}
\end{figure}
